\documentclass{bmcart}
\usepackage[utf8]{inputenc} 
\usepackage[binary-units]{siunitx}
\usepackage{multirow}
\usepackage{graphicx}
\usepackage{textgreek}
\usepackage{float}
\usepackage{tikz}		
\usetikzlibrary{fadings}
\tikzfading[name=fade out, inner color=transparent!0,outer color=transparent!100]
\tikzfading[name=fade top, bottom color=transparent!0,top color=transparent!100]
\usetikzlibrary{intersections}
\usetikzlibrary{calc}
\usetikzlibrary{positioning}
\usetikzlibrary{patterns}
\usetikzlibrary{shapes}
\usetikzlibrary{arrows}
\usetikzlibrary{decorations}
\usetikzlibrary{decorations.pathmorphing}

\usepackage{pgfplots}
\usetikzlibrary{pgfplots.patchplots}
\usetikzlibrary{pgfplots.dateplot}
\usetikzlibrary{pgfplots.colormaps}
\usetikzlibrary{pgfplots.groupplots}
\usetikzlibrary{pgfplots.polar}
\usetikzlibrary{pgfplots.units}
\usepackage{import}

\startlocaldefs
\endlocaldefs

\begin{document}

\begin{frontmatter}

\begin{fmbox}
\dochead{Research}


\title{The Bose-Einstein Condensate and Cold Atom Laboratory}


\author[
   addressref={aff1},                   
   corref={aff1},                       
   email={frye@iqo.uni-hannover.de}
]{\inits{KF}\fnm{Kai} \snm{Frye}}
\author[
addressref={aff1},
]{\inits{SA}\fnm{Sven} \snm{Abend}}
\author[
addressref={aff1},
email={bartosch@iqo.uni-hannover.de}
]{\inits{WB}\fnm{Wolfgang} \snm{Bartosch}}
\author[
addressref={aff2},
email={ahmad.bawamia@fbh-berlin.de}
]{\inits{AB}\fnm{Ahmad} \snm{Bawamia}}
\author[
   addressref={aff1},
]{\inits{DB}\fnm{Dennis} \snm{Becker}}
\author[
addressref={aff3},
email={blume@ims.uni-hannover.de}
]{\inits{CS}\fnm{Holger} \snm{Blume}}
\author[
addressref={aff4, aff5},
email={braxm@zarm.uni-bremen.de}
]{\inits{CB}\fnm{Claus} \snm{Braxmaier}}
\author[
addressref={aff6},
email={sheng-wey.chiow@jpl.nasa.gov}
]{\inits{SC}\fnm{Sheng-Wey} \snm{Chiow}}
\author[
addressref={aff7,aff8},
email={maxim.efremov@dlr.de}
]{\inits{MAE}\fnm{Maxim A.} \snm{Efremov}}
\author[
addressref={aff1},
email={ertmer@iqo.uni-hannover.de}
]{\inits{WE}\fnm{Wolfgang} \snm{Ertmer}}
\author[
addressref={aff9},
email={pfierlinger@gmail.com}
]{\inits{PF}\fnm{Peter} \snm{Fierlinger}}
\author[
addressref={aff1},
]{\inits{NG}\fnm{Naceur} \snm{Gaaloul}}
\author[
addressref={aff4, aff5},
email={jens.grosse@dlr.de}
]{\inits{JG}\fnm{Jens} \snm{Grosse}}
\author[
addressref={aff10},
email={grz@physik.hu-berlin.de}
]{\inits{CG}\fnm{Christoph} \snm{Grzeschik}}
\author[
addressref={aff11},
email={ohellmig@physnet.uni-hamburg.de}
]{\inits{OH}\fnm{Ortwin} \snm{Hellmig}}
\author[
addressref={aff2, aff10},
email={henderson@physik.hu-berlin.de}
]{\inits{VAH}\fnm{Victoria A.} \snm{Henderson}}
\author[
addressref={aff1},
email={herr@iqo.uni-hannover.de}
]{\inits{WH}\fnm{Waldemar} \snm{Herr}}
\author[
addressref={aff6},
email={Ulf.E.Israelsson@jpl.nasa.gov}
]{\inits{UI}\fnm{Ulf} \snm{Israelsson}}
\author[
addressref={aff6},
email={james.m.kohel@jpl.nasa.gov}
]{\inits{JK}\fnm{James} \snm{Kohel}}
\author[
addressref={aff2, aff10},
email={markus.krutzik@physik.hu-berlin.de}
]{\inits{MK}\fnm{Markus} \snm{Krutzik}}
\author[
addressref={aff2},
email={christian.kuerbis@fbh-berlin.de}
]{\inits{CK}\fnm{Christian} \snm{K\"{u}rbis}}
\author[
addressref={aff4},
email={claus.laemmerzahl@zarm.uni-bremen.de}
]{\inits{CL}\fnm{Claus} \snm{Lämmerzahl}}
\author[
addressref={aff12},
email={meike.list@dlr.de}
]{\inits{ML}\fnm{Meike} \snm{List}}
\author[
addressref={aff13},
email={daniel.luedtke@dlr.de}
]{\inits{DL}\fnm{Daniel} \snm{Lüdtke}}
\author[
addressref={aff14},
email={nlundbla@bates.edu}
]{\inits{NL}\fnm{Nathan} \snm{Lundblad}}
\author[
addressref={aff15},
email={jemarbur@uni-mainz.de}
]{\inits{JPM}\fnm{J. Pierre} \snm{Marburger}}
\author[
addressref={aff7},
email={matthias.meister@uni-ulm.de}
]{\inits{MM}\fnm{Matthias} \snm{Meister}}
\author[
addressref={aff15},
email={mmihm@uni-mainz.de}
]{\inits{MM}\fnm{Moritz} \snm{Mihm}}
\author[
addressref={aff16},
email={hmberkeley@gmail.com}
]{\inits{HM}\fnm{Holger} \snm{M\"{u}ller}}
\author[
addressref={aff4},
email={hauke.muentinga@zarm.uni-bremen.de}
]{\inits{HM}\fnm{Hauke} \snm{M\"{u}ntinga}}
\author[
addressref={aff3},
email={oberschulte@ims.uni-hannover.de}
]{\inits{TO}\fnm{Tim} \snm{Oberschulte}}
\author[
addressref={aff1},
email={papakonstantinou@iqo.uni-hannover.de}
]{\inits{AP}\fnm{Alexandros} \snm{Papakonstantinou}}
\author[
addressref={aff4, aff5},
email={jaka.perovsek@zarm.uni-bremen.de}
]{\inits{JP}\fnm{Jaka} \snm{Perov\u{s}ek}}
\author[
addressref={aff2, aff10},
email={achim.peters@physik.hu-berlin.de}
]{\inits{AP}\fnm{Achim} \snm{Peters}}
\author[
addressref={aff13},
email={arnau.pratisala@dlr.de}
]{\inits{AP}\fnm{Arnau} \snm{Prat}}
\author[
addressref={aff1},
email={rasel@iqo.uni-hannover.de}
]{\inits{EMR}\fnm{Ernst M.} \snm{Rasel}}
\author[
addressref={aff8},
email={albert.roura@dlr.de}
]{\inits{AR}\fnm{Albert} \snm{Roura}}
\author[
addressref={aff7, aff8, aff17},
email={wolfgang.schleich@uni-ulm.de}
]{\inits{WPS}\fnm{Wolfgang P.} \snm{Schleich}}
\author[
addressref={aff1},
]{\inits{CS}\fnm{Christian} \snm{Schubert}}
\author[
addressref={aff1, aff18},
]{\inits{STS}\fnm{Stephan T.} \snm{Seidel}}
\author[
addressref={aff13},
email={jan.sommer@dlr.de}
]{\inits{JS}\fnm{Jan} \snm{Sommer}}
\author[
addressref={aff3},
email={spindeldreier@ims.uni-hannover.de}
]{\inits{CS}\fnm{Christian} \snm{Spindeldreier}}
\author[
addressref={aff16},
email={dmsk@berkeley.edu}
]{\inits{DMSK}\fnm{Dan} \snm{Stamper-Kurn}}
\author[
addressref={aff19},
email={ben.stuhl@sdl.usu.edu}
]{\inits{BS}\fnm{Benjamin K.} \snm{Stuhl}}
\author[
addressref={aff4, aff5},
email={marvin.warner@dlr.de}
]{\inits{MW}\fnm{Marvin} \snm{Warner}}
\author[
addressref={aff1},
email={wendrich@iqo.uni-hannover.de}
]{\inits{TW}\fnm{Thijs} \snm{Wendrich}}
\author[
addressref={aff15},
email={awenzlaw@uni-mainz.de}
]{\inits{AW}\fnm{André} \snm{Wenzlawski}}
\author[
addressref={aff2},
email={andreas.wicht@fbh-berlin.de}
]{\inits{AW}\fnm{Andreas} \snm{Wicht}}
\author[
addressref={aff15},
email={windpass@uni-mainz.de}
]{\inits{PW}\fnm{Patrick} \snm{Windpassinger}}
\author[
addressref={aff6},
email={Nan.Yu@jpl.nasa.gov}
]{\inits{NY}\fnm{Nan} \snm{Yu}}
\author[
addressref={aff4, aff5},
email={lisa.woerner@zarm.uni-bremen.de}
]{\inits{LW}\fnm{Lisa} \snm{W\"{o}rner}}


\address[id=aff1]{ 
  \orgname{Institut f\"{u}r Quantenoptik, Leibniz Universit\"{a}t Hannover}, 
  \street{Welfengarten 1}, 
  \postcode{D-30167}
  \city{Hannover},
  \cny{Germany}
}

\address[id=aff2]{
  \orgname{Ferdinand-Braun-Institut, Leibniz-Institut f\"{u}r H\"{o}chstfrequenztechnik},
  \street{Gustav-Kirchhoff-Str. 4}, 
  \postcode{D-12489}
  \city{Berlin},
  \cny{Germany}    
}

\address[id=aff3]{
  \orgname{Institut f\"{u}r Mikroelektronische Systeme, Leibniz Universit\"{a}t} Hannover,
  \street{Appelstraße 4},
  \postcode{D-30167}
  \city{Hannover},
  \cny{Germany}
}

\address[id=aff4]{
  \orgname{ZARM, Universit\"{a}t Bremen},
  \street{Am Fallturm 2},
  \postcode{D-28359}
  \city{Bremen},
  \cny{Germany}    
}

\address[id=aff5]{
  \orgname{German Aerospace Center for Space Systems, DLR-RY}, 
  \street{Linzerstrasse 1},
  \postcode{D-28359}
  \city{Bremen},
  \cny{Germany}    
}

\address[id=aff6]{
  \orgname{Jet Propulsion Laboratory, California Institute of Technology},  
  \street{4800 Oak Grove Drive},
  \city{Pasadena, CA, 91109},
  \cny{USA}    
}

\address[id=aff7]{
  \orgname{Institut f\"{u}r Quantenphysik and Center for Integrated Quantum Science and Technology (IQ$^\text{ST}$), Universit\"{a}t Ulm},
  \street{Albert-Einstein-Allee 11},
  \postcode{D-89069}
  \city{Ulm},
  \cny{Germany}    
}

\address[id=aff8]{
  \orgname{Institute of Quantum Technologies, German Aerospace Center (DLR)},
  \street{S\"{o}flinger Str. 100},
  \postcode{D-89077}
  \city{Ulm},
  \cny{Germany}    
}

\address[id=aff9]{
  \orgname{Fierlinger Magnetics GmbH},
  \street{Rathausplatz 2},
  \postcode{D-85748}
  \city{Garching},
  \cny{Germany}    
}

\address[id=aff10]{
  \orgname{AG Optical Metrology, Humboldt-Universit\"{a}t zu Berlin},
  \street{Newtonstraße 15},
  \postcode{D-12489}
  \city{Berlin},
  \cny{Germany}    
}

\address[id=aff11]{
  \orgname{Institut f\"{u}r Laserphysik, Universit\"{a}t Hamburg},
  \street{Luruper Chaussee 149},
  \postcode{D-22761}
  \city{Hamburg},
  \cny{Germany}    
}

\address[id=aff12]{
  \orgname{Institute for Satellite Geodesy and Inertial Sensing, German Aerospace Center (DLR) c/o Leibniz Universität Hannover},
  \street{Welfengarten 1},
  \postcode{D-30167}
  \city{Hannover},
  \cny{Germany}
}

\address[id=aff13]{
  \orgname{Simulations- und Softwaretechnik, German Aerospace Center (DLR)},
  \street{Lilienthalpl. 7},
  \postcode{D-38108}
  \city{Braunschweig},
  \cny{Germany}    
}

\address[id=aff14]{
  \orgname{Department of Physics and Astronomy, Bates College},
  \postcode{ME 04240}
  \city{Lewiston},
  \cny{USA}    
}

\address[id=aff15]{
  \orgname{Institut f\"{u}r Physik, Johannes Gutenberg-Universit\"{a}t Mainz},
  \street{Staudingerweg 7},
  \postcode{D-55128}
  \city{Mainz},
  \cny{Germany}    
}

\address[id=aff16]{
  \orgname{Department of Physics, University of California},
  \street{366 LeConte HallMC 7300},
  \postcode{CA, 94720-7300}
  \city{Berkeley},
  \cny{USA}    
}

\address[id=aff17]{
  \orgname{Hagler Institute for Advanced Study and Department of Physics and Astronomy, Institute for Quantum Science and Engineering (IQSE), Texas A\&M AgriLife Research, Texas A\&M University},
  \postcode{TX 77843-4242}
  \city{College Station},
  \cny{USA}    
}

\address[id=aff18]{ 
  \orgname{OHB System AG}, 
  \street{Manfred-Fuchs-Str. 1}, 
  \postcode{D-82234}
  \city{Wessling},
  \cny{Germany}
}

\address[id=aff19]{
  \orgname{Space Dynamics Laboratory},
  \postcode{NM, 87106}
  \city{Albuquerque},
  \cny{USA}    
}


\begin{artnotes}
\end{artnotes}

\end{fmbox}


\begin{abstractbox}

\begin{abstract} 
Microgravity eases several constraints limiting experiments with ultracold and condensed atoms on ground.
It enables extended times of flight without suspension and eliminates the gravitational sag for trapped atoms. 
These advantages motivated numerous initiatives to adapt and operate experimental setups on microgravity platforms.
We describe the design of the payload, motivations for design choices, and capabilities of the Bose-Einstein Condensate and Cold Atom Laboratory (BECCAL), a NASA-DLR collaboration.
BECCAL builds on the heritage of previous devices operated in microgravity, features rubidium and potassium, multiple options for magnetic and optical trapping, different methods for coherent manipulation, and will offer new perspectives for experiments on quantum optics, atom optics, and atom interferometry in the unique microgravity environment on board the International Space Station.

\end{abstract}



\end{abstractbox}
%

\end{frontmatter}


\section{Introduction}

Laser cooling and the creation of Bose-Einstein condensates (BECs)~\cite{foot2005} have opened up the field of dilute ultracold quantum gases. 
These systems enable the study of fundamental aspects of quantum mechanics, such as the evolution of matter waves~\cite{RBerman1997}, the transition of quantum statistic to classical thermodynamics~\cite{Gaunt2013}, the impact of the dimensionality and topology of the gases~\cite{Sun2018} and may also probe the interface between quantum mechanics and general relativity~\cite{Schlippert2014}. 
In a typical laboratory environment gravity deforms the trapping potentials and, in the absence of magnetic or optical fields, the atoms will simply fall towards the edge of the experimental chamber limiting the available free evolution times.
Levitation techniques~\cite{Berrada2013,Leanhardt2003} can compensate gravity, but can lead to unwanted systematic shifts in experiments, for example, additional phase shifts in atom interferometers, and they become technically more complex for multiple species with different atomic masses or internal states. 

\begin{figure}[ht!]
  \includegraphics[width=0.95\textwidth]{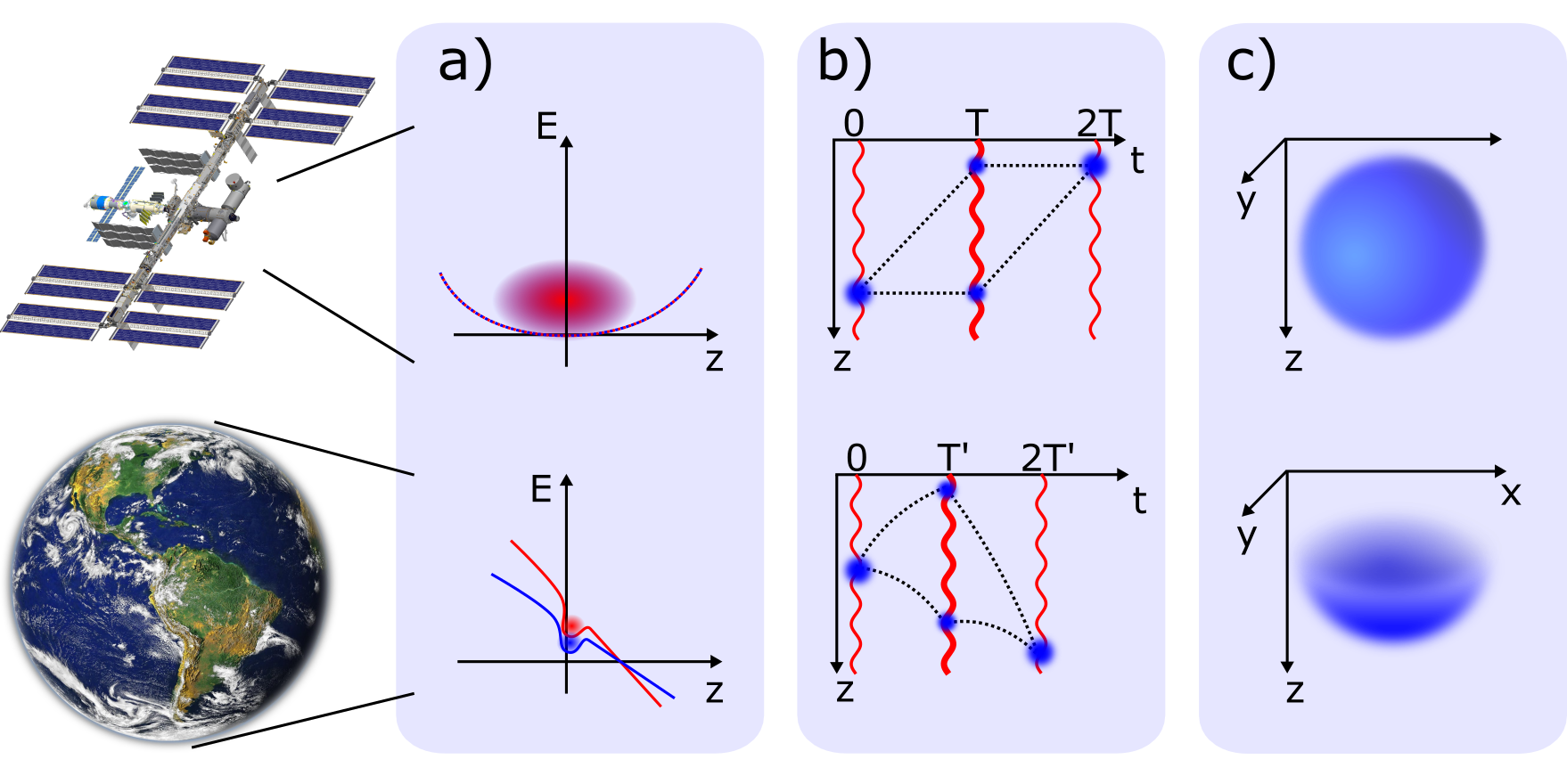}
  \caption{\csentence{Benefits of microgravity for atomic physics}
  Different experiments in microgravity (top row) and earth-bound conditions (bottom row) are shown. In all figures, the $z$-axis points in the direction of gravitational acceleration $g$. 
  \textbf{a)} Atomic species of different masses $m$ (red and blue) are confined in a potential. In the absence of any gravitational sag, the trapping potentials perfectly overlap, while in a gravitational field the two species experience a differential sag and the atomic clouds are (partially) separated. In addition, the traps have to be steeper than in microgravity to prevent the atoms from falling out of the confinement.
  \textbf{b)} Both graphs show an Mach-Zehnder atom interferometer. Laser pulses coherently split, reflect, and recombine the atomic cloud. In microgravity, the atomic trajectory is only determined by the interaction with the laser pulses and long pulse-separation times $T$ are accessible in a small setup. On ground, gravity alters the trajectory of the atoms and the free-fall distance of the atoms limits the pulse-separation time $T'$. 
  \textbf{c)} In the absence of gravitational forces new and complex trapping geometries can be realized including shell-like 3D potentials. In an earth-bound laboratory, gravity distorts such a shell trap leading to an only partially filled shell.
  }
  \label{fig:intro}
\end{figure}

In microgravity, however, the atoms stay stationary with respect to the apparatus without the need for any external holding forces, thus enabling potentials without gravitational sag, long pulse separation times in atom interferometry~\cite{RBerman1997} as well as the probing of surfaces with atoms over long times~\cite{Harber2005} and the realization of cloud geometries that are inaccessible in ground-based setups~\cite{Sun2018, Lundblad2019} as illustrated in fig.~\ref{fig:intro}.
Therefore, ultracold atoms in microgravity are ideal candidates to probe fundamental physics, such as testing the Einstein equivalence principle~\cite{Schlippert2014}, investigating the validity of quantum mechanics on macroscopic scales~\cite{Kovachy2015}, and probing dark energy~\cite{Jaffe2017} and dark matter~\cite{Graham2017, Tino2019, Elneaj2019}.
Quantum sensors based on atom interferometry are considered for earth observation with satellite gravimetry~\cite{Bidel2018}, satellite gradiometry, and navigation in space~\cite{Barrett2016c}. 
Furthermore, a flexible, microgravity cold atom system is a critical pathfinder for the demonstration of other space borne atomic physics systems, such as highly-accurate optical frequency standards.
While these applications inherently require space-borne operation, the conditions of microgravity also boost the potential performance of the proposed atom sensors.

Ultracold atom instruments have been put into microgravity using drop towers~\cite{VanZoest2010,muntinga2013,Rudolph2015,vogt2019evaporative} and sounding rockets~\cite{Becker2018} as well as with thermal atoms on board of air crafts in parabolic flights~\cite{Barrett2016d}.
Their availability, however, is limited.
The drop tower allows for only 2-3 drops per day, while sounding rockets and air crafts require extensive preparation, infrastructure, and manpower for no more than six minutes in microgravity per launch.
A novel facility, the Einstein Elevator in Hanover~\cite{Lotz2018}, is currently in the commissioning phase and will offer up to 100 shots per day, but with a limited free-fall time of approximately \SI{4}{\second}. 

The International Space Station (ISS), on the other hand, provides a permanent microgravity environment.
In May 2018 NASA launched the Cold Atom Laboratory (CAL) to the ISS~\cite{Elliott2018}, with an upgrade launched in December 2019. 
This apparatus is designed to produce ultracold degenerate quantum gases of rubidium and potassium and to perform a vast range of experiments proposed by various researchers. 
Current topics include studies on few-body dynamics, magnetic-lensing techniques~\cite{Ammann1997}, bubble-geometry traps~\cite{Sun2018}, alternative rf-outcoupling mechanisms~\cite{Meister2019}, and quantum coherence for longer than \SI{5}{\second}.

The Bose-Einstein Condensate and Cold Atom Laboratory (BECCAL), presented in this paper, is a collaboration of NASA and DLR that will serve as a multi-user and -purpose apparatus on the ISS.
It is built upon the heritage of previous and ongoing activities to offer higher atom numbers, an increased cycle rate, more complex optical and magnetic trapping strategies and improved atom-interferometry capabilities compared to the predecessor experiments.
The apparatus is designed to create BECs of $^{87}$Rb ($^{41}$K) with more than \num{1e6} (\num{1e5}) atoms.
The creation of quantum degenerate gases of $^{40}$K will also be possible.
Hence, \mbox{BECCAL} will enable the study of scalar and spinor BECs as well as of mixtures of Bose-Bose and Bose-Fermi gases.
To detect these quantum gases after several seconds of free expansion, delta-kick collimation~\cite{Ammann1997} will be employed to lower the expansion velocities of the atomic clouds.
The apparatus will support the coherent splitting of matter waves separated by several centimeters and free evolution times of several seconds.
The atoms can be exposed to blue- and red-detuned optical fields for trapping.
These fields will be spatially controllable and it will be possible to create arbitrarily-shaped potentials, which allow for versatile trapping and anti-trapping configurations.

Throughout its lifetime, BECCAL will perform a variety of experiments and serve as a pathfinder for future missions. 
This paper reports on the project, its envisioned scientific impact, and the design of the apparatus. 
The paper is organized as follows: 
Sec.~\ref{sec:SERD} lays out the scientific envelope of the mission.
The design of the instrument is detailed in sec.~\ref{sec:Design}, which includes an overview of the accommodation, budgets and safety measures (sec.~\ref{sec:overall}), the physics package (sec.~\ref{sec:PP}), the laser system (sec.~\ref{sec:LS}), the control electronics (sec.~\ref{sec:Electronics}) and the software architecture (sec.~\ref{sec:SW}).
We present the installation, operation, and maintenance procedures in sec.~\ref{sec:OPs}.
The key features of the apparatus are summarized in sec.~\ref{sec:Sum} and the conclusion is given in sec.~\ref{sec:Conclusion}.

\section{Scientific envelope}\label{sec:SERD}
Beyond principal mission targets such as the creation of BECs, the realization of low expansion velocities, and the demonstration of atom interferometry, a broader spectrum of experiments is anticipated.
Therefore, BECCAL is designed to satisfy a wide range of experimental needs.
The following sections outline several possible research topics and how they benefit from microgravity conditions.
They focus on experiments which are either not realizable on ground or which are expected to show improved performance when transferred to microgravity.
Some experiments presented here build upon and extend research that has been realized in QUANTUS, MAIUS, CAL, and other microgravity platforms.

\subsection{Atom interferometry}
Atom interferometry relies on the coherent splitting and recombination of matter waves to obtain an interference signal. 
This signal stems from the phase differences acquired on the interferometer paths.
The phase differences depend on the potentials the atoms are subjected to, and on the chosen topology~\cite{Hogan2008arXiv,Bongs2006APB,Borde2004GRG}.
Atom interferometry is a versatile tool which can be used to probe fundamental physics~\cite{Bongs2019}, such as dark matter~\cite{Arvanitaki2018} and dark energy theories~\cite{Jaffe2017}, and to test the universality of free fall (UFF)~\cite{Schlippert2014,Rosi2017NatComm,Barrett2016NatComm,Duan2016PRL,Zhou2015,Bonnin2013PRA,Tarallo2014PRL}.
Furthermore, it is possible to test the validity of quantum mechanics in extreme parameter ranges by creating coherent superpositions spanning over several tens of centimeters~\cite{Kovachy2015} or interferometer times larger than \SI{10}{s}. 
Atom interferometry also provides sensitive and accurate sensors for measuring Earth's gravitational acceleration~\cite{Bidel2018,Freier2016,Louchet2011} for geodetic observations and offers new perspectives for navigation~\cite{Barrett2016c}. 

In a microgravity environment, the free fall of the atoms does not generally limit the total interferometer time. 
This allows for space and time separations of coherently split atoms which, on ground, are solely realizable in long baseline facilities~\cite{Hartwig2015, Kovachy2015, Zhou2015}.
Extended free-fall times can increase the sensitivity of atom interferometers.
For the common Mach-Zehnder like~\cite{MarkKasevichandStevenChu1991} interferometer scheme, the phase scales with the squared pulse separation time.
Microgravity also enables the operation of an atom interferometer that is stationary with respect to nearby objects in order to probe surface effects or to realize novel interferometry methods.
A nadir pointing sensitive axis allows for measurement campaigns for mapping Earth's gravitational field~\cite{Trimeche2019,Douch2017a, carraz2014spaceborne} and UFF tests~\cite{Tino2013NuclPhysB,Aguilera14CQG,Williams16NJP}.
Furthermore, these activities serve as a pathfinder for future missions and applications, e.g. deep-space inertial sensing with atom interferometry, and as a gateway towards precision measurements of gravity, measurements of Newton's gravitational constant~\cite{Rosi2014Nature}, the gravitational Aharanov-Bohm effect~\cite{Hohensee2012PRL} and gravitational wave detection~\cite{Hogan2016PRA,Hogan2011GRG,Dimopoulos2008PRD}. 

\subsection{Coherent atom optics}
An experiment in microgravity will likely exploit delta-kick-collimation (DKC) techniques~\cite{Ammann1997} to create BECs with expansion velocities smaller than \SI{100}{\micro \meter\per\second}, which makes long free evolution times within the restricted volume possible.
The lack of gravitational acceleration enables constant, highly controllable velocities of the atoms, which correspond to nearly monochromatic matter waves.
These matter waves will be manipulated with various spatial- and temporal-dependent light fields to study basic phenomena of linear optics, but with switched roles -- the matter acts as the wave and the light acts as the dispersive or absorptive medium.
The latter depends on whether the scattering process is chosen to be coherent or incoherent and enables situations where light sheets diffract and refract matter waves.

For instance, a laser beam blue-detuned with respect to the atomic $D_2$-transition creates a repulsive potential and reflects the matter waves like a mirror.
Two atom mirrors opposite each other form a matter-wave resonator.
Microgravity conditions allow for full 3D symmetry of such a resonator so that atoms can be trapped in a 3D box potential, closely reproducing the textbook example of a particle in a box with infinitely high potential walls. 

Another relevant topic is atom optics in the non-linear regime where homogeneous magnetic fields are applied to tune the scattering length of the atoms with the help of Feshbach resonances and to control the dispersion.
For example, a negative scattering length creates a situation analogous to self-focusing.
In addition, coherent matter-wave mixing will be possible, which enables a large set of new interferometric configurations. 

Quantum reflection on surfaces represents another subject of interest.
Here, a microgravity environment can help to improve the existing results~\cite{Pasquini2006} of quantum-mechanical reflection of matter-waves  from solid surfaces by providing smaller incident velocities and reduced mean-field interaction effects thanks to the shallow trapping possible in microgravity.

\subsection{Scalar Bose-Einstein condensates}
Scalar BECs will be studied in various magnetic and optical potentials, where the absence of gravity allows the creation of extremely shallow traps with trapping frequencies well below \SI{1}{Hz} whilst maintaining atomic trapping.
Since the critical temperature of Bose-Einstein condensation in a harmonic potential is proportional to the geometrical mean of the trapping frequencies, shallower traps result in a lower critical temperature. 
The critical temperature is the upper temperature limit of a BEC and lowering this threshold means that one can assure colder condensates~\cite{Leanhardt2003}.
This approach also leads to a decrease of the entropy of the gas, unlike adiabatic decompression techniques and DKC. 
The creation of spin excitations inside a BEC~\cite{Olf2015} further reduces the entropy and the momentum distribution of the thermalized magnons gives an estimate of the entropy. 
Hence, the creation of a gas with a record-low entropy per particle of  $S/N \leq\SI{1e-4}{k_B}$ appears to be within reach in microgravity.
Low-entropy gases are an essential resource for probing subtle many-body quantum effects and enhance the fidelity of quantum reflection on surfaces. 

Magnetically trapped atoms can be transferred to an untrapped state via a weak radio-frequency field.
In microgravity the dynamics of the outcoupled atoms are governed only by the repulsive atom-atom interaction giving rise to a slow expanding shell of atoms. 
This so-called space atom laser features an almost spherically-symmetric shape even if the initial BEC was confined in an elongated trap~\cite{Meister2019}. 

Another closely related topic is 3D bubble shells of trapped BECs~\cite{Sun2018, Lundblad2019}.
Here, a strong radio-frequency field dresses the trapping potential and hollows out the BEC.
In this way a 2D BEC can be created embedded in a closed 3D shell.
This process is solely realizable in microgravity and enables studies on thermodynamics and collective modes at a dimensional cross-over, edgeless surface excitations, and topological constraints on vortex dynamics in a curved geometry. 

\subsection{Spinor Bose-Einstein condensates and quantum gas mixtures}
Spinor Bose gases are optically-trapped BECs where the spin of the atoms can be changed without loosing the confinement. 
This is in contrast to magnetically trapped condensates, where the spin needs to be polarized properly to form a weak-field seeking state. 
Since the first realization of an optical dipole trap~\cite{Chu1986, Stamper-Kurn1998}, spinor Bose gases have been the subject of extensive research due to their interplay of magnetism and super-fluidity~\cite{Baranov2008, Stamper-Kurn2013}.
In microgravity, the differential gravitational sag vanishes for different elements, isotopes and states enabling studies of spinor BECs and mixtures in unprecedented regimes. 

For instance, optical box potentials with edge lengths of up to \SI{100}{\micro\meter} and particle numbers within reach of BECCAL lead to particle densities in the order of \SI{1e12}{\centi\meter^{-3}}.
This allows the study of magnetic interactions in a quantum fluid, such as spin-dependent {\it s}-wave collisions and spin-mixing dynamics. 
Increasing the volume of the trap lowers the densities of the atomic gases and allows the exploration of the non-interacting regime where the atoms are expected to show long-lived spin coherence.
Thus, such an experiment has potential applications as a magnetic and rotation sensor, in addition to being a probe of fundamental physics. 

Moreover, homogeneous magnetic fields enable the tuning of inter- and intra-species scattering lengths.
The magnetic field strength can be chosen to enable the creation of strongly interacting mixtures independent of the trap volume~\cite{Roberts2000}. 
Several Feshbach resonances may be accessible with reasonable magnetic field strengths to create hetero-nuclear spinor gases.

Furthermore, the combination of intercomponent attraction and intracomponent repulsion allows the formation of self-bound quantum gases, so-called quantum droplets~\cite{Petrov2015, Cabrera2018, Semeghini2018}. 
These systems have never been studied in microgravity, which would enable the observation of the long-term dynamics of such droplets in a potential-free environment.

\subsection{Strongly interacting gases and molecules}
Strong atom-atom interactions often enhance three-body processes and fuel rapid atom loss. 
Microgravity increases the realizable interaction times due to the accessibility of extremely low densities and temperatures.
In particular, the time-of-flight before the application of a magnetic lens determines the size of the atomic cloud and therefore controls its density. 
Additionally, the interaction volume can be chosen to be potential-free (e.g. a box trap generated by blue-detuned light) to create a homogeneous interaction strength over a large volume.
Moreover, clouds of different atomic species perfectly overlap resulting in higher rates and efficiencies of Feshbach-molecule formation~\cite{DIncao2017}. 

Ultra-low temperatures offer the energy resolution required for observing new features of Efimov physics~\cite{Braaten2006, Naidon2017}. 
In addition, tuning the scattering rate with magnetic fields allows one to access a new regime, where two-body collisions are negligible with respect to three-body collisions, but without a high loss rate due to three-body recombination.

\subsection{Quantum information}
Dedicated experiments on quantum optics and atom optics may serve as a pathfinder and a major technology demonstrator for advancing quantum technology into space. 
This may enable future satellite-based quantum cryptography and quantum information/state transfer between distant points on Earth or beyond. 
In particular, cold-atomic ensembles could be useful for quantum information processing~\cite{Northup2014, Lvovsky2009} for space-based quantum communications applications~\cite{Bedington2017, Liao2018}.
BECCAL's flexible cold atom platform will allow for demonstrations of techniques such as electromagnetically-induced transparency or coherent population trapping in ultracold ensembles for topics such as optical quantum memories and slow light~\cite{Ma2017}.

\section{Instrument design}\label{sec:Design}

In the following section the instrument design is presented, starting with an overview of the system, the restrictions set by the operation in orbit, and the requirements for safety. 
Subsequently the individual subsystems are explained in more detail. 

\subsection{Overview} \label{sec:overall}

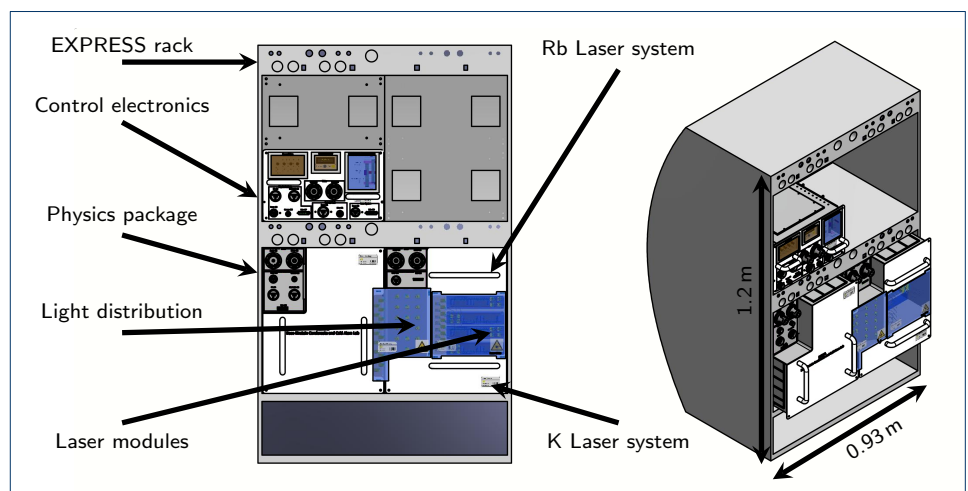
\begin{figure}[ht]
	  \begin{tikzpicture}[>=stealth]
	\tikzset{every node}=[font=\footnotesize\sffamily]
	\node[inner sep=0pt] (Picture) at (0,0)
	{\includegraphics[width=0.95\textwidth]{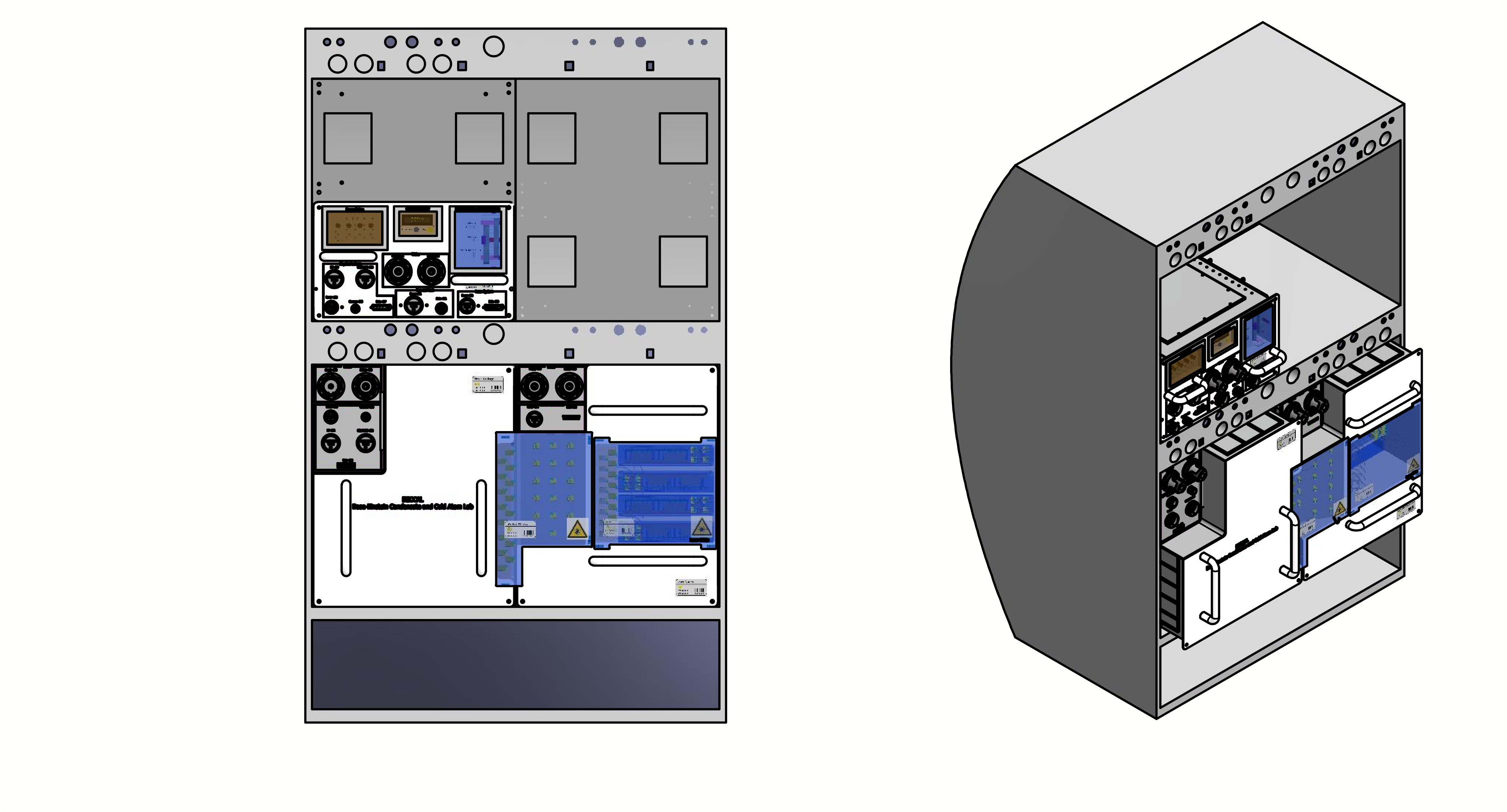}};
	
	\def\rend{5.4}
	\def\expr{-3.6}
	\def\lend{-5.4}

	\draw[<-,line width=2pt](\expr,2.8)--(\lend,2.8)node[right,above]{EXPRESS rack};
	\draw[<-,line width=2pt](\expr,1)--(\lend,2)node[right,above]{Control electronics};
	
	\draw[<-,line width=2pt](-0.5,0)--(1.2,2.7)node[left,above]{Rb Laser system};
	\draw[<-,line width=2pt](-0.5,-1.5)--(1.2,-2)node[left,below]{K Laser system};
	\draw[<-,line width=2pt](\expr,0)--(\lend,0.5)node[right,above]{Physics package};
	\draw[<-,line width=2pt](-1.5,-0.7)--(\lend,-0.8)node[right,above]{Light distribution};
	\draw[<-,line width=2pt](-0.5,-0.8)--(\lend,-2)node[right,below]{Laser modules};
	\draw[<->,line width=2pt](3.25,-2.75)--(5.3,-1.5)node[pos=0.6, below, sloped]{\SI{0.93}{\meter}};
	\draw[<->,line width=2pt](3.1,-2.55)--(3.1,1.3)node[pos=0.6, above, sloped]{\SI{1.2}{\meter}};

	\end{tikzpicture}
  \caption{\csentence{Overview of the apparatus}
      The available space for BECCAL in the EXPRESS rack is shown. 
      Cables and fibers are omitted for clarity.
      One single locker contains the control electronics, one double locker the laser system, and another double locker the physics package.
      The front panels will have handles for the astronauts to hold on and to ease the installation.
      The light produced by the laser system is guided via optical fibers to the physics package.
      The fibers are protected by an interlocked cover. 
      }\label{fig:EXPRESS_Rack}
      \end{figure}

BECCAL will be housed in the Destiny module on the ISS. 
The space station provides a standardized rack system to house payloads of different size and purpose, the so-called EXPRESS racks (EXpedite the PRocessing of Experiments to Space Station).
Each EXPRESS rack offers eight standardized compartments and
NASA allocates five of eight lockers for BECCAL (fig.~\ref{fig:EXPRESS_Rack}). 
For implementation and due to launch restrictions, these five lockers will be separated into one single and two double lockers. 
The remaining locker spaces of the EXPRESS rack are occupied by other payloads. 
The accommodation in an EXPRESS rack therefore sets stringent requirements on volume, mass, external power, thermal management (sec.~\ref{sec:SWaP}), emitted radiation, and safety (sec.~\ref{sec:Safety}). 

The payload is divided into three subsystems -- the physics package, the laser system, and the control electronics.
The physics package, which contains the ultra-high vacuum system, where all the experiments on the atoms are carried out, is located in one double locker (sec.~\ref{sec:PP}).
The other double locker contains the laser system and the laser electronics (sec.~\ref{sec:LS}). 
The single locker will house parts of the control electronics and the on-board computer (sec.~\ref{sec:Electronics}). 
The computer will run the experiment control software (sec.~\ref{sec:SW}) and will have additional hard drives for experimental data storage.

\subsubsection{SWaP budgets}\label{sec:SWaP} 

This section summarizes the available and chosen budgets for size, weight, and power (SWaP). 
Additionally, the thermal control system is outlined. 

Compared to conventional quantum-optics-laboratory experiments, volume, mass, and power consumption of the apparatus are significantly reduced.
The single locker has a size of \SI{66}{\liter} and the two double lockers take \SI{164}{\liter} each.
The total mass is limited by the EXPRESS rack specifications to \SI{328}{\kilo \gram}.

The external power consumption depends on the operation mode and on the details of the experimental sequence.
For more details on the power consumption see sec.~\ref{sec:Electronics}. An overview of the different operational modes is given in sec.~\ref{sec:OPs}. 

During operation, the payload generates thermal loads that will be dumped to the air cooling system of the EXPRESS rack and the ISS medium temperature water cooling loop.
BECCAL's inner structure blocks the airflow through the system, which decreases the efficiency of the thermal transport, thus, most of the thermal load will be dissipated through the shared water cooling system.
The maximum thermal load that can be dissipated is therefore limited to comply with ISS interface requirements.
In BECCAL, each subsystem features two quick couplers and a piping system that enables water cooling of the installed heatsinks.

\subsubsection{Safety}\label{sec:Safety}
BECCAL is operated on board the ISS, and consequently in an environment designed for human space flight. 
While this offers unique opportunities for access, maintenance, manipulation, operation, and exchange of payloads, the safety and security of all parties involved, especially on orbit, is crucial. 
The safety requirements cover any risks for humans (i.e. by excess heat or radiation), especially for ISS crew members. 
Moreover, the safety of the space station has to be ensured and any interference (e.g. by electromagnetic radiation or vibrations) with other payloads needs to be minimized. 
Consequently, mitigating and preventing potential hazards and hazardous situations are design drivers for BECCAL. 
In particular, high voltage lines and free-space laser-light paths are contained within three levels of containment, such as interlocked containers and female sockets on the payload's front plates. 

BECCAL is operated from ground and does not require crew interaction in nominal operation. 
Even during installation and exchange operations, the lockers are not opened by the crew, who can only access the interfaces between the systems. 
The lockers are designed to prevent potential internal faults from propagating to the outside, this includes structural resilience, counter measures to block sharp objects entering the habitable volume, shielding against radiation (including light fields), and safeguards against excess surface temperatures. 

Electromagnetic interference of the experiment with the electrical devices on board the ISS potentially disturbs the experiment as well as the operation of the space station. 
The passive magnetic shielding (see also sec.~\ref{sec:PP}), which is more commonly used to shield atomic physics experiments from external magnetic fields~\cite{Kubelka-Lange2016}, also protects the ISS from the magnetic fields and radio-frequency radiation produced by the apparatus.
The level of stray fields and radiation will be verified by a standard electromagnetic radiation test.

Like CAL, BECCAL will use rubidium and potassium as atomic sources. 
These elements are very reactive, radioactive with half lives of $>$\SI{1e9}{} years, and can be poisonous if swallowed or inhaled.
However, the small amount of material used, and the enclosure inside several layers of protection, mitigates the radiation exposure of the crew and surroundings. 
Additionally, both alkali metals condense when exposed to ambient air pressures, reducing the probability of the release of large quantities of particles into the habitable area in the event of a breach in the vacuum system.  

The requirements on radiation hardness against cosmic radiation are moderate compared to deep space missions due to the ISS's low orbit of \SIrange{370}{460}{\kilo\meter}. 
Only single event effects need to be taken into account.
Critical parts of the electronics will be radiation hard and shielded. 
All other circuits are designed to not go into over-current, short circuit or over-temperature, if a single component fails. 

Temperature sensors are distributed all over the experiment, not only to detect over-temperature, but also to characterize any warm-up effects and the temperature distribution inside the payload~\cite{Haslinger2018}.

In addition to these passive mitigation strategies, the payload will be surveyed from ground during experimental operation. 
In standby the system will constantly check for key parameters, such as vacuum quality, temperature levels, and drawn power, and alert the payload developer team and ISS personnel if a parameter exceeds the acceptable range for nominal operation. 
In general, the payload will be shut down by the ground personnel before a hazardous situation arises, as a fault would damage the system before endangering a crew member or the ISS.  

\subsection{Physics package}\label{sec:PP}

\begin{figure}[!ht]
    \begin{tikzpicture}[>=stealth]
  \tikzset{every node}=[font=\footnotesize\sffamily]
  \node[inner sep=0pt] (Picture) at (0,0)
  {\includegraphics[width=0.95\textwidth,keepaspectratio]{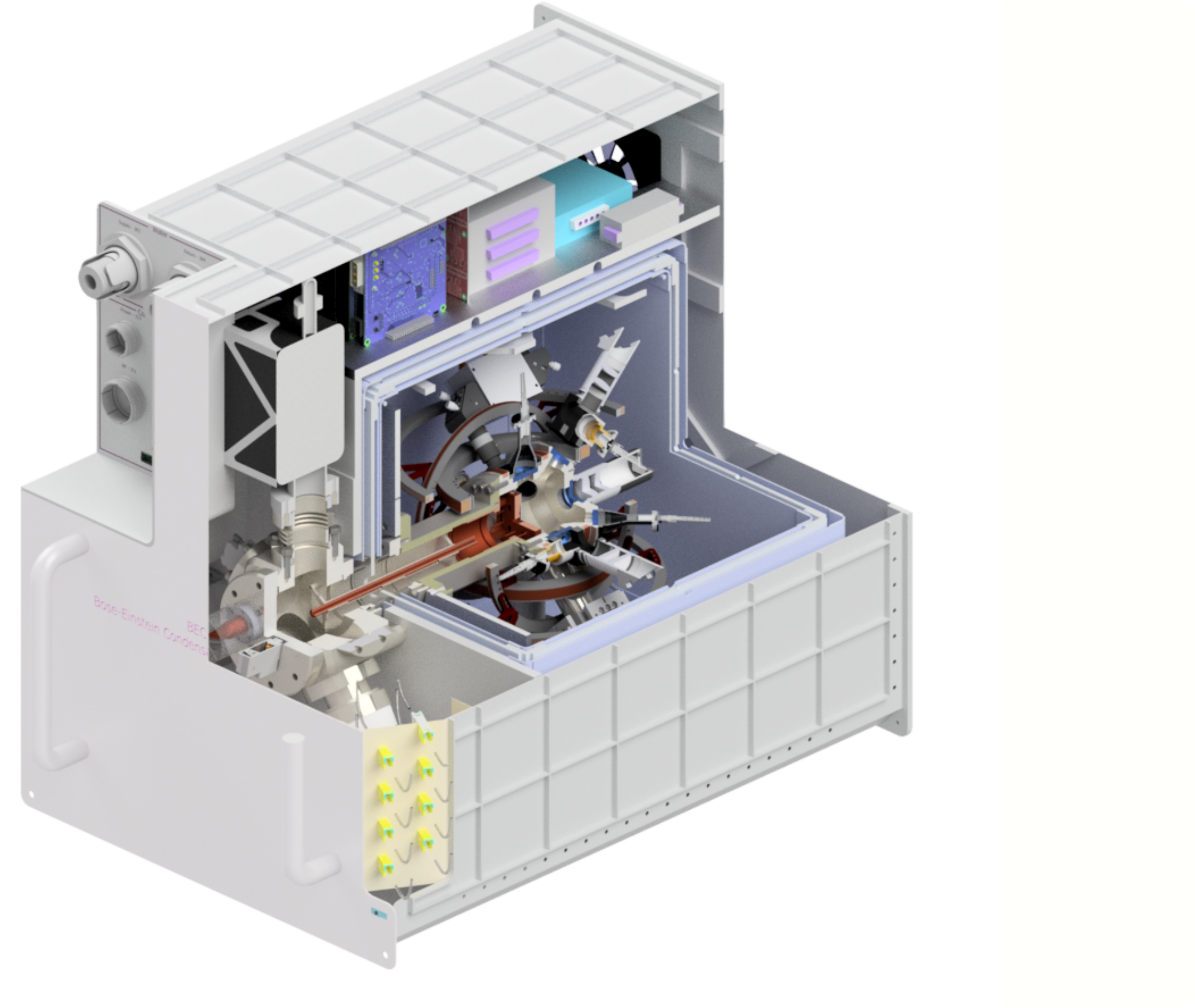}};
  
  \def\rend{4}
  
  \draw[<-,line width=2pt](0,1)--(\rend,1)node[left,above]{Science Chamber};
  \draw[<-,line width=2pt](0,3)--(\rend,3)node[left,above]{Physics Package electronics};
  \draw[<-,line width=2pt](2.1,-.5)--(\rend+1,-0.5)node[left,above]{Magnetic shield};
  \draw[<-,line width=2pt](-3,-1.5)--(\rend+1,-1.5)node[left,above]{Pump System};
  \draw[<-,line width=2pt](-2.1,-3.5)--(\rend,-3.5)node[left,above]{Fiber Patch Panel};
  
  \end{tikzpicture}
  \caption{\csentence{Cross section through the physics package}
      The front panel, including handles, is shown on the left-hand side of the picture.
      The fiber connectors interface the physics package to the fibers coming from the laser system.
      The interior of the locker is not crew-accessible.
      An ion getter pump (IGP), a cold cathode gauge vacuum sensor and the electrical feed-throughs for the atom chip are located directly behind the front panel.
      Other parts of the vacuum system are encapsulated in a magnetic shield. 
      The shield consists of two mu-metal layers and an aluminium layer in between.
      A custom made titanium sublimation pump is located inside the shield to be close to the science chamber.
      This ultra-high vacuum chamber is pumped through holes inside the copper mounting structure for the atom chip. 
      The atom chip sits in the center of the science chamber. 
      Three sets of coils attached to the chamber create homogeneous magnetic fields. 
      One additional pair creates magnetic fields inside the chamber with a strength of up to \SI{175}{G}. 
      The physics package locker also contains parts of the control electronics. 
      In particular, controllers and drivers with high current and high voltage output are placed in this locker to shorten the transmission paths. 
      All cables and fibers are omitted for clarity.
      }
\end{figure}

\noindent
The physics package is the subsystem where the experiments on the atoms are performed.
It contains the ultra-high vacuum system, which consists of several connected vacuum chambers, which are adapted from QUANTUS~\cite{Rudolph2015} and MAIUS setups~\cite{Becker2018}.
In the preparation chamber, atomic gases of rubidium and potassium are cooled to form a cold atomic beam, which points into the science chamber (sec.~\ref{sec:ScienceCh}).
This chamber supports trapping, cooling, and manipulation of atomic clouds (sec. \ref{sec:Atomic_source}).
The two-chamber design allows for evaporation, preparation, and transport of the atoms while simultaneously maintaining the ultra-high vacuum in the science chamber. 
The science chamber is surrounded by wire coils, which produce magnetic fields (sec.~\ref{sec:MagField}) in combination with the atom chip.
Atom interferometry can be performed in two orthogonal axes (sec.~\ref{sec:AI}).
Also attached to the science chamber are the setup for the creation of arbitrary shaped optical potentials (sec.~\ref{Sec:PaintPot}) and the detection system (sec.~\ref{Sec:Detection}).
The vacuum pressure inside the chambers is maintained by several pumps, which are described in sec.~\ref{sec:Vacuum}.

\subsubsection{Science chamber}\label{sec:ScienceCh}

\begin{figure}[!ht]
\begin{tikzpicture}[>=stealth]
    \tikzset{every node}=[font=\footnotesize\sffamily]
  \node[inner sep=0pt] (Picture) at (0,0)
  {\includegraphics[width=0.95\textwidth,keepaspectratio]{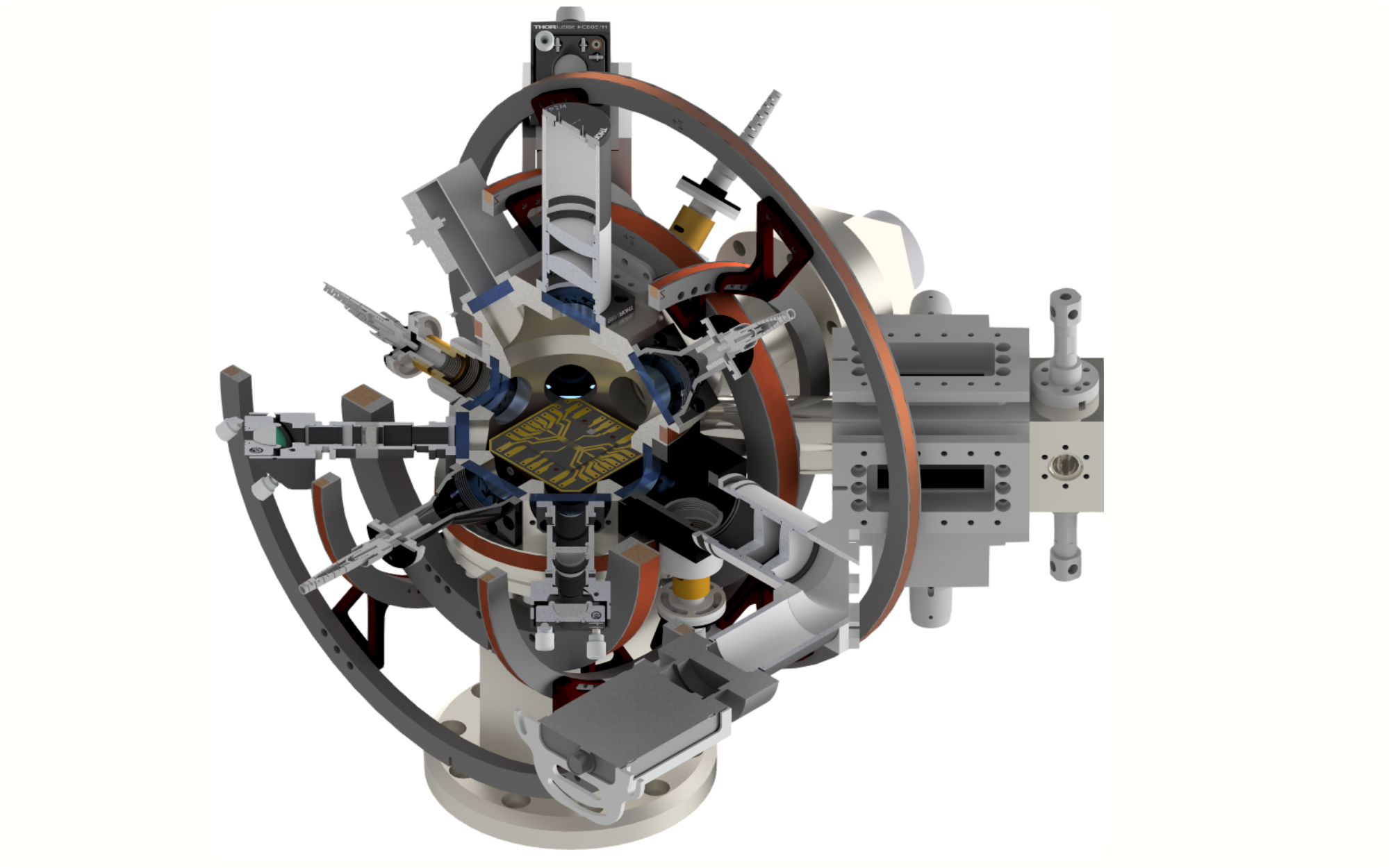}};
  
  \def\rend{4.5}
  \def\lend{-4.5}
  
  \draw[<-,line width=2pt](1.3,2)--(\rend-2,3)node[left,above]{Piezo tip-tilt stage};
  \draw[<-,line width=2pt](1.8,1.5)--(\rend-0.5,2)node[left,above]{Preparation chamber};
  \draw[<-,line width=2pt](3.3,1)--(\rend+0.5,1)node[left,above]{Oven};
  \draw[<-,line width=2pt](3.5,-1.1)--(\rend,1);
  \draw[<-,line width=2pt](0,-2.5)--(\rend-2,-2.5)node[right,above]{Detection};
  
  \draw[<-,line width=2pt](-3,1)--(\lend,2)node[right,above]{MOT telescopes};
  \draw[<-,line width=2pt](-1,0)--(\lend-1,1)node[right,above]{Atom chip};
  
  \draw[<-,line width=2pt](-2.5,-2.5)--(\lend,-2.5)node[right,above]{Helmholtz coils};

\end{tikzpicture}

  \caption{\csentence{Three quarter cross section through the science chamber}
        The science chamber is manufactured from a single piece of titanium alloy (Ti6Al4V) and has an octagonal shape with an inner diameter of \SI{6.5}{\centi \meter}. The top part, which opposes the atom chip, provides five additional viewports. In total, eleven viewports around the chamber allow for optical access. The atom chip sits in the center of the chamber and the whole setup is surrounded by four pairs of copper-wire coils. 
        All cables and fibers are omitted for clarity and the y-axis points along gravity. 
}\label{fig:sciencechamber}
\end{figure}

\noindent
The science chamber has an octagonal shape and is manufactured from a titanium alloy (Ti6Al4V).
The chamber features several ports for optical access and vacuum connectors - a total of six viewports and two vacuum connections as well as a top part with five additional viewports.
It contains an atom chip and the tip-tilt mirror used for atom interferometry (sec. \ref{sec:AI}).

Four viewports are available as optical access for magneto-optical trap (MOT) beams and for the absorption detection system (sec. \ref{Sec:Detection}).  
Two beams are reflected by the surface of the atom chip.
Two viewports perpendicular to each other will be used for the creation of painted, blue-detuned optical potentials (sec. \ref{Sec:PaintPot}).
The light for a red-detuned crossed optical dipole trap will be transmitted through two viewports.
The two ports facing nadir are used for atom interferometry.
One of those two ports provides an extension to the vacuum system to house the tip-tilt stage, and the other contains a window through which the atom interferometry beams enter the chamber.
All the optical components are directly mounted onto the chamber to provide the required pointing stability.
The fiber collimators contain additional photodiodes for power and polarization monitoring. 

\subsubsection{Atomic source}\label{sec:Atomic_source}
This section describes the creation of ultracold atomic clouds of rubidium and potassium.
First, we comment on the choice of elements.
Then we lay out the route for rapid production of quantum degenerate gases. 

Rubidium and potassium were chosen for several reasons: (i) The strong $D_2$ transition lines of rubidium and potassium are only \SI{13}{\nano\meter} apart. 
An optical broadband coating with that range is easy to manufacture and is found in many commercial off-the-shelf optics. 
Thus, the same optics can be used for both elements which reduces the complexity of the apparatus substantially.
(ii) The interspecies interaction can be tuned by using the rich spectrum of Feshbach resonances at moderate magnetic field strengths. 
(iii) There is roughly a factor of two in mass difference which makes the combination of those two elements interesting for potential tests of the universality of free fall.
(iv) Potassium offers bosonic and fermionic isotopes, which facilitates the study of Bose-Bose and Bose-Fermi gases of different species.
(v) Both species are currently being used in the MAIUS-B experiment, which features a similar atom chip, as well as in the CAL experiment.
BECCAL will profit from this experience.

The cooling of the atoms takes place in two chambers.
A $\mathrm{2D}^+$-MOT configuration~\cite{Chaudhuri2006} creates a cold atomic beam in the preparation chamber.
This beam will load a 3D-magneto-optical trap in the science chamber through a differential pumping stage.
We expect the flux of $^{85}$Rb and $^{87}$Rb into the 3D-MOT to be \SI{1e9}{atoms \per \second} each, which was already shown in a similar setup~\cite{Rudolph2015}.
The flux for $^{39}$K ($^{41}$K) is expected to be roughly a factor of 10 (100) lower than for rubidium. 
The preparation chamber is directly connected to the ovens.
One oven contains \SI{1}{g} rubidium, the other \SI{1}{g} potassium, both in natural abundance.
An additional dispenser will contain \SI{100}{\milli \gram} of enriched $^{40}$K.
These samples are sufficient to last the whole mission duration. 

Together with the magnetic field, four light beams form the 3D-MOT.
Two light beams are reflected by the coated surface of the atom chip with an incident angle of \ang{45}.
We expect the MOT to provide \num{2e9} $^{87}$Rb atoms, \num{1e9} $^{85}$Rb atoms, \num{8e8} $^{39}$K atoms, \num{4e8} $^{41}$K atoms and \num{1e7} $^{40}$K atoms for single species operation each.
The subsequent cooling steps are discussed in detail in Ref.~\cite{Rudolph2015}, while here, a short outline is given.
The 3D-MOT is followed by a phase of compressed MOT, with lower magnetic fields and higher detuning.
This results in a higher density than in the initial MOT. 
The magnetic fields are then switched off and the cloud is cooled with polarization-gradient cooling. 
A shallow Ioffe-Pritchard magnetic trap with large trapping volume captures the atoms and subsequently the magnetic field gradients are increased.
This tight trap allows for fast evaporation as a consequence of the increased collision rate in the dense cloud.
The evaporation efficiency is such that there will be more than three orders of magnitude gain in phase-space density for every order of magnitude in atom loss.
We expect to create single-species Bose-Einstein condensates of $^{87}$Rb and $^{41}$K
with more than \num{1e6} and \num{1e5} atoms, respectively. 
For the creation of a dual-species BEC, the atoms are loaded into an optical dipole trap after a short period of evaporation.
The Feshbach field is switched on to reduce two- and three-body losses and optical evaporation is performed. 
For mixtures the atom numbers are expected to be one order of magnitude less than for single species BECs. 
Magnetic trapping of the fermionic isotope of potassium will be realized for at least \num{1e4} atoms with a Fermi fraction $(k_b T/E_f)$ of $\leq1$.

For further manipulation, the atoms can be transferred into various magnetic traps (sec. \ref{sec:MagField}) and optical traps (sec. \ref{Sec:PaintPot}).
The atoms can be subjected to an magnetic~\cite{Ammann1997} or optical delta-kick collimation to reduce the expansion velocity to values $\leq$ \SI{100}{\micro \meter \per\second}.

\subsubsection{Atom chip and magnetic fields}\label{sec:MagField}

The magnetic fields are created by an atom chip inside the science chamber and ex-vacuo coils in Helmholtz configuration.
The atom chip is a versatile and compact tool for generating magnetic fields~\cite{Rudolph2015, Berrada2013}.
Its design is adapted from the QUANTUS-2 apparatus~\cite{Rudolph2015}.
It consists of one layer with mesoscopic copper wire structures and two layers of microscopic wire structures.
The atom chip will feature \SI{1}{\milli\meter\squared} surfaces off-center with different electrical properties to study quantum reflection.
Since the atoms are only a couple of hundreds of micrometers to a few millimeters away from the atom chip surface, high gradients can be realized with lower power than with a coil setup outside the vacuum chamber. 
High trap frequencies (up to \SI{2}{\kilo\hertz}) can be achieved with a few amperes of current, which makes evaporative cooling fast and power efficient.
Most structures on the chip are made from single wires, which results in much lower inductance than in a coil setup. 
This makes switching times shorter than \SI{100}{\micro\second} possible. 

In order to create the quadrupole fields required to operate a MOT or Ioffe-Pritchard-type magnetic traps, the field created by the atom chip itself needs to be overlapped with homogeneous bias fields.
In this setup, the homogeneous fields are created by three Helmholtz coils outside the vacuum chamber; one pair for every direction.
The center of the harmonic traps can be chosen to be \SIrange{0.1}{2}{\milli\meter} away from the chip surface with aspect ratios of $100\!:\!100\!:\!1$ to $2\!:\!2\!:\!1$.
An additional set of four coils produces homogeneous fields of up to \SI{175}{G} for tuning the scattering length of the atoms. 
This field can be modulated with an on-top amplitude of \SIrange{0}{1}{G} and frequencies up to \SI{100}{\hertz}, to enable fast hopping into or across Feshbach resonances.

The atom chip will have additional structures for radio-frequency and microwave fields. 
These fields will be used for forced evaporative cooling, state preparation, and to create the dressed potentials for ring or bubble shaped traps.
In addition to the micro structures on the atom chip, there will be microwave antennas attached to the chamber. These antennas provide homogeneous fields required for large volume bubble-shaped traps.

The science and the preparation chambers and the coils are located within a three-layer magnetic shield: two layers of mu-metal with a layer of aluminum in between. 
Mu-metal is a soft-ferromagnetic alloy commonly used for magnetic shielding due to its high permeability.
The magnetic shielding factor will be roughly 300 for static fields and several orders of magnitude higher for oscillating fields.

\subsubsection{Atom interferometry setup}\label{sec:AI}
BECCAL will have two independent atom interferometry axes. 
The primary axis is parallel both to Earth's acceleration (nadir) and the surface of the atom chip.
The beam will be reflected by a mirror inside the vacuum apparatus.
The second interferometry axis is perpendicular to the first and the beam is reflected by the atom chip. 
The $1/e^2$ beam waist will be \SI{6}{mm} for both axes.

Auxiliary sensors will be used to characterize the environment and for correlation with the atom interferometers.
These comprise an accelerometer and a gyro sensor which will be placed in the vicinity of the vacuum system.
The science chamber, and in particular the atom chip, will be equipped with a grid of temperature sensors.
The temperature data will be used to model acceleration due to black body radiation~\cite{Haslinger2018}.

The primary interferometer axis is parallel to the earth's gravitational acceleration, which enables the determination of the differential Eötvös ratio for rubidium and potassium~\cite{Schlippert2014}. 
The light beam enters the chamber via a window, interacts with the atoms, then passes through a quarter-waveplate, is retro-reflected by a mirror and passes through the same waveplate again before interacting with the atoms again.
The waveplate rotates the parallel linear polarization of the two incident light fields by \ang{180}, to enable double-Raman diffraction beam splitters~\cite{DoubleDiffRaman}.

Wavefront distortions are one of the limiting factors for the accuracy of atom interferometers~\cite{Freier2016,Louchet2011}.
The distortions introduced to the light before interacting with the atoms are common to both the incident and counter-propagating beams.
They are therefore suppressed in most atom interferometry configurations. 
However, this is not the case for distortions introduced by optics behind the atoms in the optical path.
To minimize this, the retro-reflecting mirror and the waveplate will be designed to have a global flatness of $\lambda/20$ and a local root mean squared surface roughness of $\lambda/100$ and will be placed inside the vacuum chamber.

\begin{figure}
\includegraphics[width=0.45\textwidth]{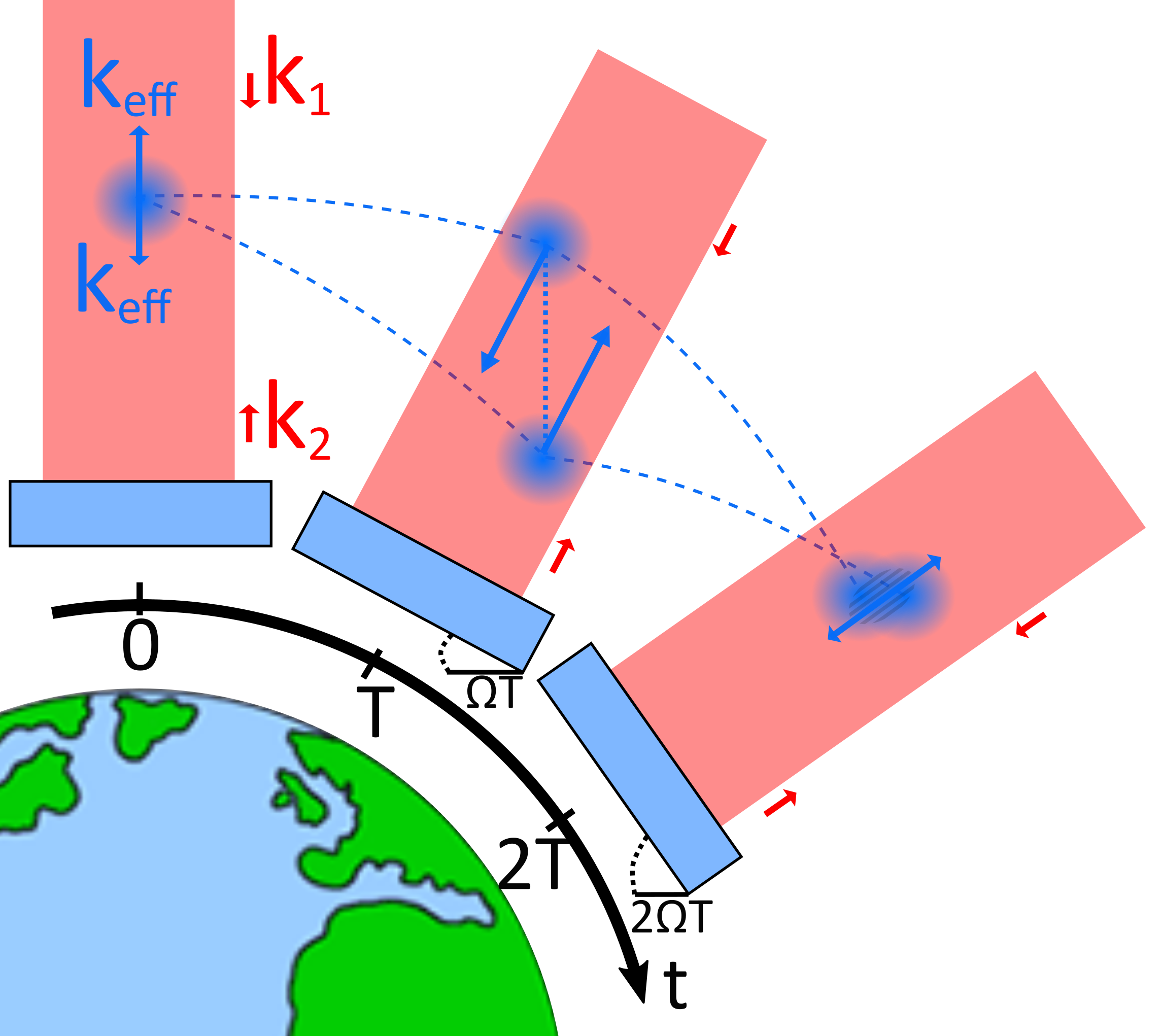}
  \includegraphics[width=0.45\textwidth]{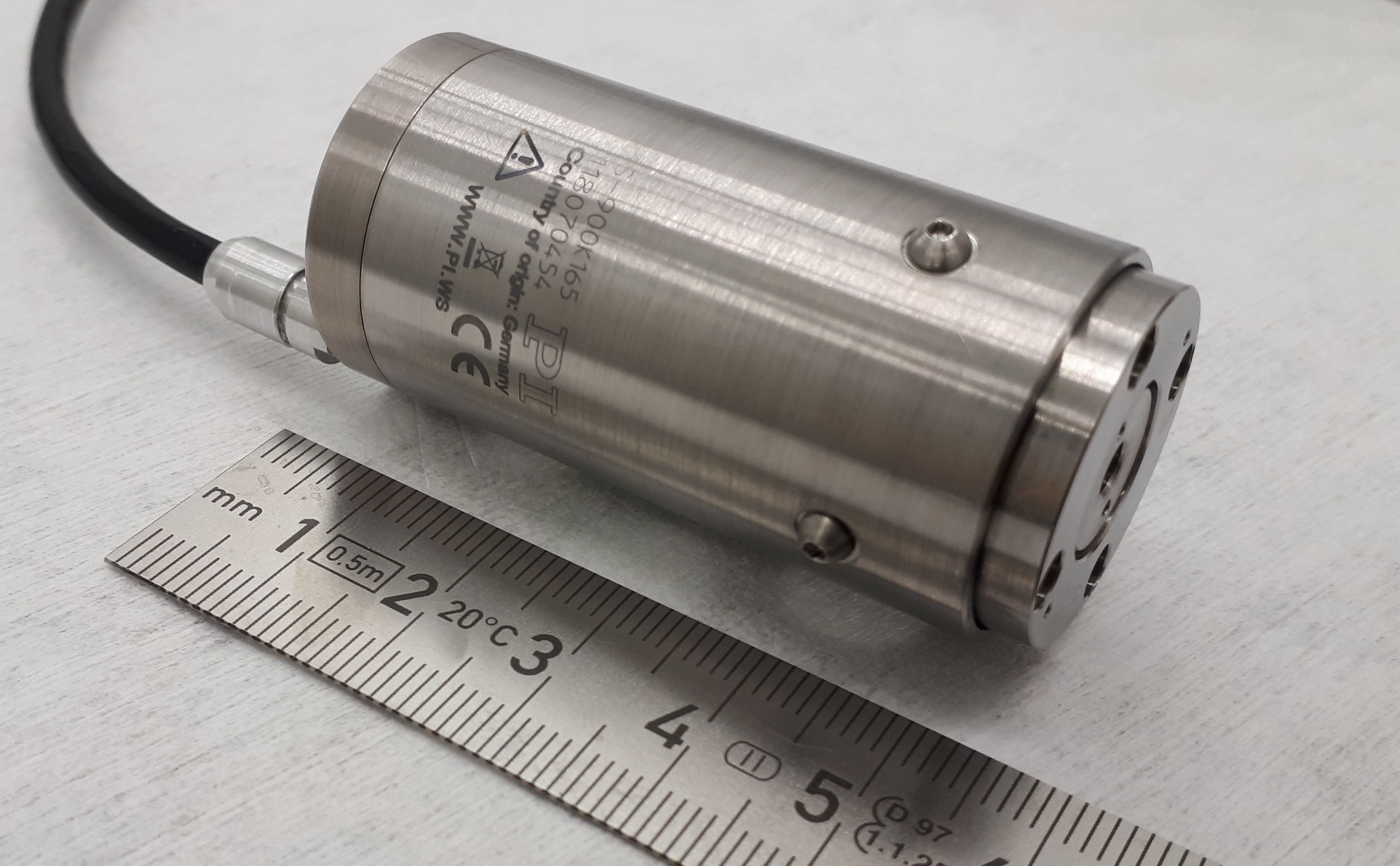}
  \caption{\csentence{Atom interferometry in a rotating environment}
    Left: Schematic of a double-diffraction atom interferometry sequence~\cite{DoubleDiffRaman} in a rotating environment with angular velocity $\Omega$. The picture shows different snapshots of the sequence at three different times. The atoms (blue) are coherently split at the time $t=0$ by a retro-reflected light beam (red) with wave numbers $k_1$ and $k_2$. The dotted blue line depicts the trajectory of the atoms. The atomic momentum $k_\mathrm{eff}=k_1-k_2$ is reversed at $t=T$ and another splitting pulse is applied at $t=2T$. The rotation causes the clouds to not perfectly overlap, which reduces the contrast. Right: The tip-tilt stage used for rotation compensation. The titanium cylinder has a \SI{25.4}{\milli\meter} diameter and a length of \SI{53.5}{\milli\meter}, and contains four piezo actuators. They move the mirror with a full stroke of $\pm$\SI{2}{\milli \radian}. To reduce positioning inaccuracies, strain gauges are applied to the piezo actuators. This allows for closed-loop operation of the stage, giving lower positioning inaccuracies for long times without re-calibration.
      }\label{fig:PTT}
\end{figure}

The ISS orbits the earth with an angular velocity of $\Omega=\SI{1.1}{\milli \radian\per\second}$.
Therefore, without any counter-rotating device for the interferometry mirror, the orbiting motion of the ISS will eliminate the contrast of the interferometry fringes completely (fig. \ref{fig:PTT}). 
While the systematic errors can in principle be accounted for in post-correction, the loss of contrast due to rotation is irretrievable.
The contrast for a pulse separation time of \SI{900}{\milli\second} will be below \SI{0.2}{\percent}~\cite{Roura2014} when assuming an atom interferometer using a $^{87}$Rb BEC with a mean velocity spread of \SI{100}{\micro\meter\per\second} and a Thomas-Fermi radius of \SI{150}{\micro \meter}.
It has been shown that counter-rotating the interferometry mirror against external rotations is an effective method to regain contrast~\cite{Lan2012}.
To this end, in our setup, the interferometry mirror sits on top of a moving stage driven by piezo actuators.
Fig. \ref{fig:PTT} (right) shows the custom design created in collaboration with \textit{Physik Instrumente}.
The rotating mirror will compensate the rotation of the ISS for the required \SI{2.6}{\second} total interferometry time.

The compensation of the rotation with a tip-tilt mirror implies a change in the effective wavenumber between the subsequent atom-light interactions of an interferometer pulse sequence. Fortunately, the combination of gravity gradients and these residual rotational effects can be effectively mitigated with the technique proposed in Ref.~\cite{Roura2017} and experimentally demonstrated in Ref.~\cite{Overstreet2018}. For total interferometer times of $2 T = \SI{2.6}{\second}$ this involves changing the single-photon frequencies by $\Delta\nu  \approx \SI{1.5}{\giga\hertz}$ for the mirror pulse in a Mach-Zehnder interferometer~\cite{Roura2017b}.

\subsubsection{Optical potentials}\label{Sec:PaintPot}

The apparatus will support experiments requiring light fields red- or blue-detuned to the $D_2$ transition, with arbitrary shaping possible with the blue-detuned optical potentials.
Blue-detuned potentials enable trapping the atoms in the dark region of a geometry, which results in a homogeneous trap~\cite{Gaunt2013} with low atom-light scattering rates~\cite{Franke-Arnold2007}.

Two options for the creation of highly controllable optical potentials are investigated for use in BECCAL. 
One option is the use of 2D acousto-optical deflectors (AODs), which deflect a beam of light in any transverse direction with an adjustable angle.
By steering the beam much faster than the trap-frequency of the atomic trap, the atoms will see a time-averaged potential~\cite{Henderson2009, Trypogeorgos2013, Bell2016}.
The refreshing rate of every possible shape shall be above \SI{100}{\hertz} in order to produce a time-averaged static potentials for the atoms. 
Another option is the use of spatial light modulators (SLMs).
They imprint a phase onto a light beam to change the intensity distribution at a certain point in space~\cite{Bruce2011, Lee2014}. 
With both options, it will be possible to create blue detuned optical traps with a (time-averaged) barrier height of \SI{5}{\micro \kelvin} for contour lengths of $\leq$\SI{100}{\micro\meter}.
The inner trap size will be tunable from \SIrange{20}{1500}{\micro\meter}.
Also the optical barrier can be moved with a speed of up to \SI{5}{\milli\meter\per\second}.

The red-detuned light is used to create a crossed optical dipole trap~\cite{Chu1986, Stamper-Kurn1998}. 
One beam of this trap can be steered with an AOD to ensure the overlap of the beams. 
The center of this trap is located \SI{2}{\milli \meter} away from the atom chip surface.
The focal point will have a waist of \SI{100}{\micro \meter} and the potential depth is tunable from \SIrange{0.01}{5}{\micro \kelvin}. 
Deep optical dipole traps are not required because the trap does not need to support the atoms against gravity. 
It will be possible to add a retro-reflection of one beam to create an optical lattice prior to every experimental run.

\subsubsection{Detection}\label{Sec:Detection}

Virtually every experimental sequence in BECCAL will conclude by measuring an atomic signal.
The apparatus will feature two acquisition methods.
The first is a fluorescence detection system based on collecting photons that the atoms emit after an excitation.
We utilize a system of multiple lenses with a large numerical aperture for $2f-2f$ imaging with $f$ being the focal length of the lens system.
This creates an image of the atoms on a photo diode. 
The scheme uses near-resonant light and is state- and species-dependent.
It will be used to obtain loading and lifetime curves of the atomic clouds as well as a state-selective readout for interferometry~\cite{Schlippert2014}.

The second detection method relies on the atomic absorption of photons from an illuminating laser beam. 
A camera will take images of the laser beam with and without the shadow of the atomic cloud to gain information of the cloud's density. 
Two perpendicular absorption imaging systems will provide 3D information of the size and the shape of the atomic cloud.
Each absorption detection system consists of a collimator to provide the illumination light.
On the opposite side of the chamber is a lens systems which will focus the absorption images of the atoms onto a camera sensor.
In addition to static lenses, we will utilize electrically tunable lenses to have a variable focal length $f$.
In principle, the imaging system also resembles a $2f-2f$ scheme, but with two lenses.
The first lens is optimized to minimize spherical aberrations and sits $2f$ away from the atoms.
The second lens sits close to the camera and is electrically tunable. 
For this purpose, we utilize a shape-changing lens from \textit{optotune} with an optical fluid, which is sealed off with an elastic polymer membrane.
This lens changes the focal length of the imaging systems by \SI{\pm 5}{\milli \meter} and allows us to image the atoms in different positions with a resolution $\leq$\SI{10}{\micro\meter}. Additionally, the depth of field will be adjustable from \SIrange{100}{2000}{\micro\meter}.
This is especially useful for quantum reflection experiments as the different surfaces are off-center with respect to the atom chip.
Furthermore, the investigation of hollow BECs will profit from this feature since the focus can be shifted to the surface of the shell.

\subsubsection{Infrastructure}\label{sec:Vacuum}

The science chamber will be continuously pumped by an ion getter pump (IGP) and a passive titanium sublimation pump. 
The combination of these pumps allows for the effective pumping of different species and maintains a vacuum quality in the order of \SI{1e-10}{mbar} throughout the operation of BECCAL. 
To maintain the vacuum during times of standby, the IGP is operational at all times. 
At regular intervals, and to increase the vacuum quality should it be needed, the titanium sublimation pump will be activated. 

As the IGP produces strong magnetic fields, it is mounted outside the magnetic shield alongside an access port for the cables, and a vacuum sensor. 
The entirety of this system is called the pumping cube. 
The pumping cube is also the access point for the roughing pump, which is installed during integration to establish the required vacuum quality for the IGP. 
As the roughing pump is both large in volume and high in mass it is not transported to the ISS but pinched off before launch.

The launcher will not provide electrical power.
Consequently, the active vacuum pump is switched off during this period of highest mechanical load, i.e. integration into the launcher, launch, and transport to the ISS.   
In combination with the absence of the roughing pump in orbit, a strong limit is set on this unpowered duration.
In order to allow for the re-activation of the IGP in orbit, the vacuum system is designed to keep the pressure below \SI{1e-5}{mbar}, even under the conditions described above.

Control electronics for the atom chip, the coils, the high voltage power supply for the IGP, and auxiliary vibration, rotation and magnetic field sensors are also located outside of the magnetic shield.
The controller for the IGP sits close to the pump so that the high voltage line is very short.

Electrical connectors are embedded in the outer locker case to provide connections for the control electronics subsystem and for the scientific data to be sent from this subsystem to the data storage device. 
There are also seventeeen optical fiber connectors which will be used to connect the laser system and the physics package.

\subsection{Laser system}\label{sec:LS}

All light fields required for BECCAL are generated and controlled within the `laser system' subsystem. 
This system (including its associated control electronics) occupies one EXPRESS rack double locker. 
The generated light is delivered to the physics package via polarization-maintaining single-mode optical fibers. 
Within this section, we will outline the capabilities of the laser system, the architecture, and describe the main technologies used within the laser system, including the lasers, the free-space optics and the fiber optics.

\subsubsection{Laser system capabilities}

The laser system must deliver light to seventeen separate optical fibers to the physics package in order to facilitate the ambitious functionality of BECCAL as laid out in sec.~\ref{sec:SERD} and~\ref{sec:PP}.  Each of these light paths must be controlled in terms of power, frequency, and polarization, as well as having dynamic control on suitable timescales.

The choice of elements has been motivated in sec.~\ref{sec:Atomic_source}. 
Additionally, in the context of the laser system, the D${}_2$ transitions of Rb and K at \SI{780}{\nano\metre} and \SI{767}{\nano\meter}, respectively, are easily accessible and have compatible natural linewidths with the light emitted by diode lasers~\cite{Wieman1991}.

The required light specifications are summarized in tab.~\ref{tab:lsreq} and described below.

\begin{table}[t!]
	\caption{Overview of the light fields delivered to the physics package from the laser system. The given power values correspond to the total power per set of fibers. All power values are estimates of the required and achievable powers rather than a guarantee of the power delivered.}
	\begin{tabular}{lcccc}
		\hline
		\\
		\textbf{Function} & \textbf{Number of} & \textbf{Wavelength} & \multicolumn{2}{c}{\textbf{Power (\si{mW})}} \\
		& \textbf{connections} & \textbf{(\si{nm})} & \textbf{Cool} & \textbf{Repump} \\
		\\
		\hline
		\\
		\multirow{2}{*}{Detection} & \multirow{2}{*}{2} & \num{780} & \num{6} & \num{2} \\
		& & \num{767} & \num{7} & \num{7} \\
		\multirow{2}{*}{3D-MOT} & \multirow{2}{*}{4} & \num{780} & \num{90} & \num{12} \\
		& & \num{767} & \num{75} & \num{65} \\
		\multirow{2}{*}{2D-MOT} & \multirow{2}{*}{4} & \num{780} & \num{80} & \num{40} \\
		& & \num{767} & \num{70} & \num{70} \\
		\multirow{2}{*}{Interferometry 1} & \multirow{2}{*}{1} & \num{780} & \num{15} & \num{15} \\
		& & \num{767} & \num{15} & \num{15} \\
		\multirow{2}{*}{Interferometry 2} & \multirow{2}{*}{2} & \num{780} & \num{70} & \num{70} \\
		& & \num{767} & \num{70} & \num{70} \\
		Dipole trap & 2 & \num{1064} & \multicolumn{2}{c}{\num{300}} \\
		Blue box & 2 & $<$\num{767} & \multicolumn{2}{c}{\num{50}} \\
		\\
		\hline
	\end{tabular}\label{tab:lsreq}
\end{table} 

In sec.~\ref{sec:Atomic_source} the MOT arrangement was described in terms of the physics package. 
In order to facilitate this, the laser system provides four light paths for the \mbox{2D$^+$-MOT}: two MOT beams, a pushing beam, and a retarding beam. In each of these paths, cooling and repumping light for both potassium and rubidium are overlapped. For the 3D-MOT, again four paths are provided for MOT beams with all four necessary frequency components overlapped in each fiber. It is possible to detune the light frequency by up to approximately $\pm$\SI{200}{\mega\hertz} from the cooling and repumping transitions.  In addition to producing a 3D-MOT, these beams will be used for sub-Doppler cooling prior to evaporation, thus necessitating the ability to smoothly ramp both frequency and intensity in these paths. 

Two dipole trap paths at \SI{1064}{\nano\meter} are provided for optical evaporation and other techniques such as DKC.

As outlined in sec.~\ref{sec:AI}, BECCAL will have two independent atom interferometry axes with light driving Raman transitions. 
Three separate paths deliver light to the two interferometry axes.
One fiber will provide light for the primary interferometry axis.
For the second interferometry axis, two individual paths are provided, as single Raman diffraction is to be realized on that axis. In the primary axis, two frequency components per species are overlapped in the fiber, whereas for the secondary axis, one component per species is used per fiber. The two different frequency components per species have a relative detuning equal to the ground state hyperfine splitting of the relevant isotope plus the two-photon recoil frequency and a variable detuning. 

Two detection paths are provided for the techniques discussed in sec.~\ref{Sec:Detection}. Each of these paths have nominally the same power of all four frequency components. These light paths will also be used to provide light pulses suitable for quantum optics and information experiments and as such it will be possible to have a single frequency of light per fiber, with a different frequency component in each fiber.

The final two light paths provide the light required for blue-detuned dipole potentials. 
This light will have a wavelength below \SI{767}{\nano\meter}.

All lasers except those used for dipole potentials (both $<$~\SI{767}{\nano\meter} and \SI{1064}{\nano\meter}) must be locked to an atomic reference transition. We are able to use either modulation transfer spectroscopy (MTS) or frequency modulation spectroscopy (FMS) to lock two reference lasers -- one locked to a transition in K and the other to a transition in Rb.  
All remaining lasers are offset-locked to the two reference lasers. 
Further details of this scheme can be found in sec.~\ref{sec:ZD}.

Dynamic control of optical power will be implemented with three possible regimes: fast switching times of less than \SI{10}{\micro\second} to \SI{-30}{\decibel}, slower complete extinction in less than \SI{10}{\milli\second}, and linear modification across the dynamic range in \SI{1}{\milli\second}. 
This applies to all light paths delivered to the physics package except the interferometry beams, for which fast switching must be possible in less than \SI{1}{\micro\second}. 
Optical power will be stabilized and measured before each experimental run.

\subsubsection{Laser system architecture}\label{sec:ls_ar}

\begin{figure} [!t]
	\begin{tikzpicture}[>=stealth]
	\tikzset{every node}=[font=\footnotesize\sffamily]
	\node[inner sep=0pt] (Picture) at (0,0)
	{\includegraphics[width=0.95\textwidth,keepaspectratio]{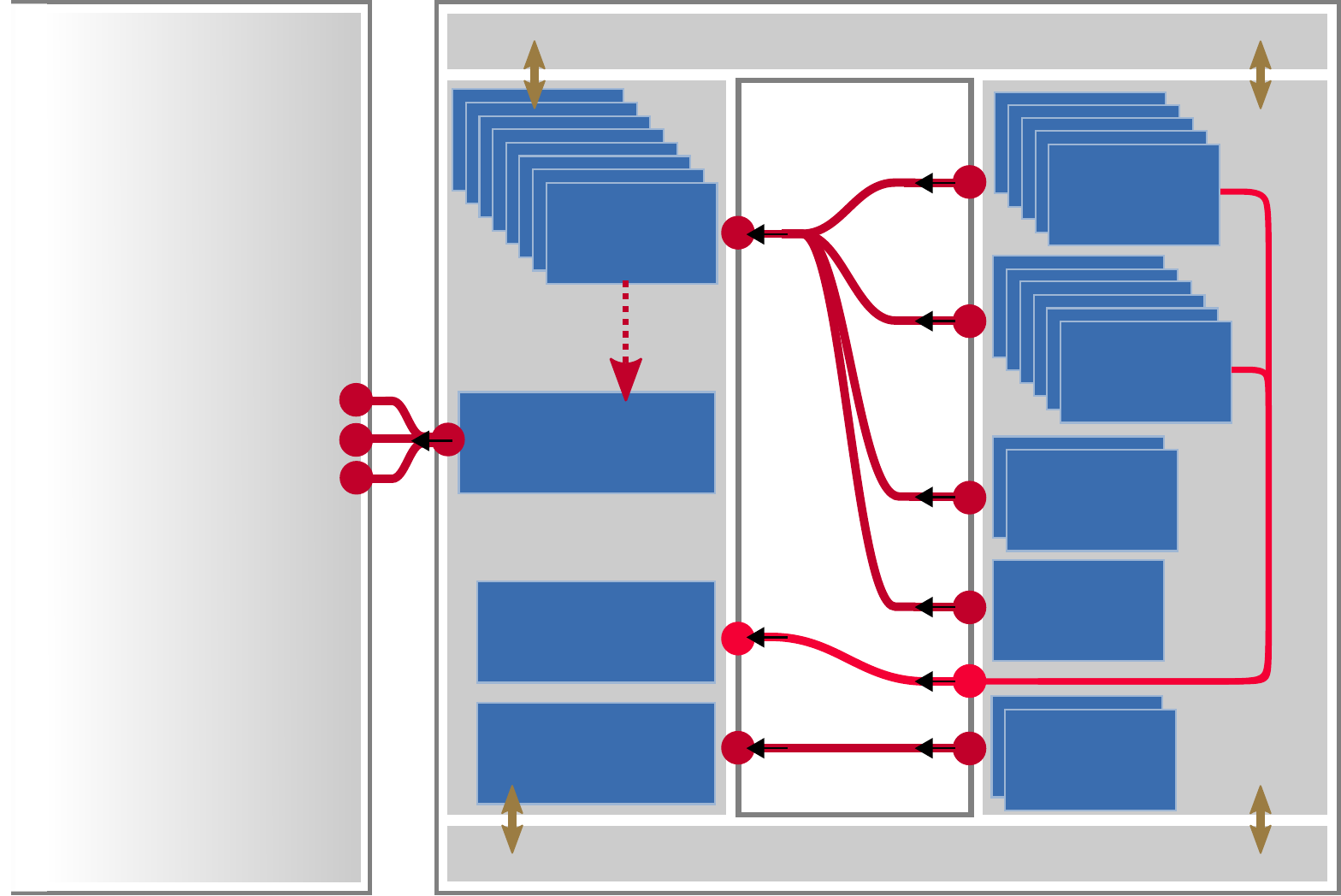}};
	
	\def\rend{4.5}
	\def\lend{-4.5}
	\definecolor{mylb}{rgb}{0.620, 0.713, 0.831}
	\definecolor{mydb}{rgb}{0.027, 0.145, 0.349}
	\node[align=center, text width=2cm, text=mylb] at (-.35,1.9) {8 Optical benches};
	\node[align=center, text width=1cm, text=mylb] at (4.15,2.3) {Lasers \num{5} $\times$ Rb};
	\node[align=center, text width=1cm, text=mylb] at (4.25,.65) {Lasers \num{6} $\times$ K};
	\node[align=center, text width=1cm, text=mylb] at (3.75,-.43) {\SI{1064}{\nano\meter} Lasers};
	\node[align=center, text width=1cm, text=mylb] at (3.63,-1.47) {$<$\SI{767}{\nano\meter} Laser};
	\node[align=center, text width=1cm, text=mylb] at (3.7,-2.8) {Reference lasers};
	\node[align=center, text width=2cm, text=mylb] at (-.75,0.05) {Fiber-based distribution};
	\node[align=center, text width=2cm, text=mylb] at (-.65,-1.67) {Offset locking};
	\node[align=center, text width=2cm, text=mylb] at (-.65,-2.75) {Spectroscopy bench};
	
	\node[align=center, text=mydb] at (1.65,3.65) {Rb laser control electronics};
	\node[align=center, text=mydb] at (1.65,-3.65) {K laser control electronics};
	\node[align=right, text=mydb] at (-4.25,0.05) {Detection $\times$ \num{2}\\
	2D-MOT $\times$ \num{4}\\
	3D-MOT $\times$ \num{4}\\
	Interferometry $\times$ \num{3}\\
	Red dipole $\times$ \num{2}\\
	Blue dipole $\times$ \num{2}};

	\node[align=center, rotate=270, text=mydb] at (5.7,0.05) {Lasers housed in ORUs};
	\node[align=center, rotate=90, text=mydb] at (-5.65,0) {Physics package};
	\node[align=center, rotate=90] at (.9,0.1) {Crew accessible area};
	\node[align=center, rotate=90] at (-2.4,-2) {Crew accessible area};
	
	\end{tikzpicture}

	\caption{\csentence{A functional block diagram of the laser system}
	Block diagram showing the approximate physical layout of the laser system and the light paths through the system. Inter-locker light carrying fibers are shown in red, with each line representing multiple fibers, and each circle representing multiple connectors. The lighter red indicates low optical power fibers used for locking We note that each inter-locker fiber connection is a 1-to-1 cable. Control signals from the electronics are shown in gold.  
	} \label{fig:LSBlock}
\end{figure}

The laser system is organized into three subsystems: lasers, free-space distribution, and fiber-based distribution. A functional block diagram showing these subsystems is shown in fig.~\ref{fig:LSBlock}.

To deliver the optical powers shown in tab.~\ref{tab:lsreq}, a total of sixteen lasers are used. 
Six lasers emit light with a wavelength of \SI{780}{\nano\meter} to address Rb (including one reference laser), seven lasers emit light with \SI{767}{\nano\meter} to address K (including one reference laser), two lasers for the creation of red-detuned dipole traps at \SI{1064}{\nano\meter}, and one laser runs at a wavelength below \SI{767}{\nano\meter} for blue-detuned optical potentials. 
The light is created within an extended cavity diode laser master oscillator power amplifiers (ECDL-MOPA) configuration. Further details of these lasers are given in sec.~\ref{sec:lasers}.  
They are shown on the right hand side of fig.~\ref{fig:LSBlock}.  
The sixteen lasers will be housed in four removable boxes of four lasers each which are called orbital replaceable units (ORUs). 
This ORU based approach is taken as a form of redundancy such that a subset of lasers can be swapped out of the payload with minimal disturbance in the event of degradation.

The constraints set by the SWaP budget (sec.~\ref{sec:SWaP}) and the complexity of the experiment favor the use of free-space optics rather than fiber-based optical components.
We integrate optics on ultra-stable optical benches made of Zerodur. 
These boards are presented in sec.~\ref{sec:ZD}. 
This technology is also used in the form of a spectroscopy bench in order to lock our lasers to an atomic reference.  

\begin{figure}
	\begin{tikzpicture}[>=stealth]
	\tikzset{every node}=[font=\footnotesize\sffamily]
	\node[inner sep=0pt] (Picture) at (0,0)
	{\includegraphics[width=0.95\textwidth,keepaspectratio]{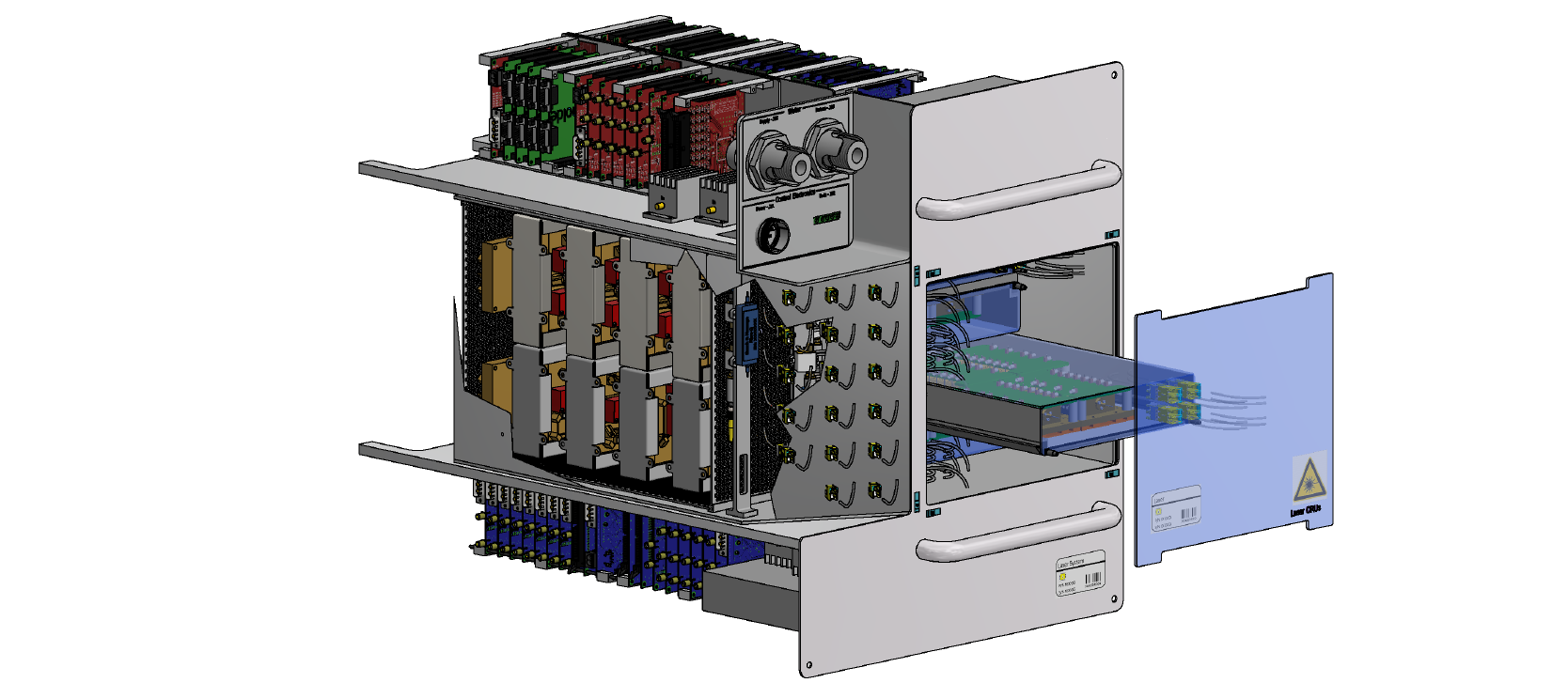}};
		
  \def\rend{4.5}
\def\lend{-4.5}

\draw[<-,line width=2pt](2.2,0)--(\rend,1)node[left,above]{Laser ORU};
\draw[<-,line width=2pt](3.5,-1.6)--(\rend-.5,-2)node[left,below]{Interlocked Panel};
\draw[<-,line width=2pt](-2.5,2)--(\lend,2)node[right,above]{Rb Laser Electronics};
\draw[<-,line width=2pt](-2.6,-.5)--(\lend,-.5)node[right,above]{Zerodur Boards};
\draw[<-,line width=2pt](-2.6,.5)--(\lend,.5)node[right,above]{Fiber Optics};
\draw[<-,line width=2pt](-2.5,-1.5)--(\lend,-1.5)node[right,above]{K Laser Electronics};

	\end{tikzpicture}
	\caption{\csentence{Overview of LS CAD}
	A preliminary CAD drawing of the laser system locker. The riveted structure contains electronics at the top and bottom, with orbital replaceable units containing lasers on the right, free space benches sitting on the left of the locker, and fiber optics and rf-based items will be housed in the remaining space.
} \label{fig:LSCADOverview}
\end{figure}

Many of the components of the laser system have been qualified through their use in sounding rocket and drop tower missions~\cite{Schkolnik2017, Schkolnik2016, Pahl2019, Doringshoff2019,  Lezius2016, Dinkelaker2017}.  For example, the current generation of laser modules flew in the JOKARUS mission~\cite{Schkolnik2017, Doringshoff2019}.
Furthermore, similar laser modules, Zerodur-based optical benches and commercial fiber-based components flew in MAIUS-1 \cite{Schkolnik2016}, FOKUS \cite{Lezius2016} and KALEXUS \cite{Dinkelaker2017}.  In addition to flight-based qualification, ground based testing as well as vibration and shock tests have been carried out on components~\cite{Wicht2017, Schiemangk2015, Mihm2019}. 

The safety requirements of the ISS (as described in sec.~\ref{sec:Safety}) are of particular relevance to the laser system. 
As a result, the laser system is entirely inaccessible to the crew with the exception of the interface between the laser system and the physics package (seventeen optical connections), and between the laser ORUs and the rest of the laser system (up to thirty-two optical connections). 
Both sections are protected by interlocked panels.

Due to the large number of optical connections to be crew-made, it is necessary to chose fiber connectors which are both intuitive to use and robust over multiple mating cycles. 
The fiber connections from the laser system to the physics package will be made during installation, whereas the ORU connections will be made prior to flight (or by crew in the event of replacement).
Each optical connection involves an optical fiber patch cord in a crew-accessible part of the payload and two mating sleeves (interfacing with the internal parts of the payload).  

In order to ensure minimal optical losses, maximal polarization stability, and ease of use, \textit{E2000} connectors and mating sleeves are being tested in a laboratory environment and the necessary additional qualification tests are planned.

\subsubsection{Lasers} \label{sec:lasers}

\begin{figure}
			\begin{tikzpicture}[>=stealth]
	\tikzset{every node}=[font=\footnotesize\sffamily]
	\node[inner sep=0pt] (Picture) at (0,0)
	{\includegraphics[width=0.95\textwidth,keepaspectratio]{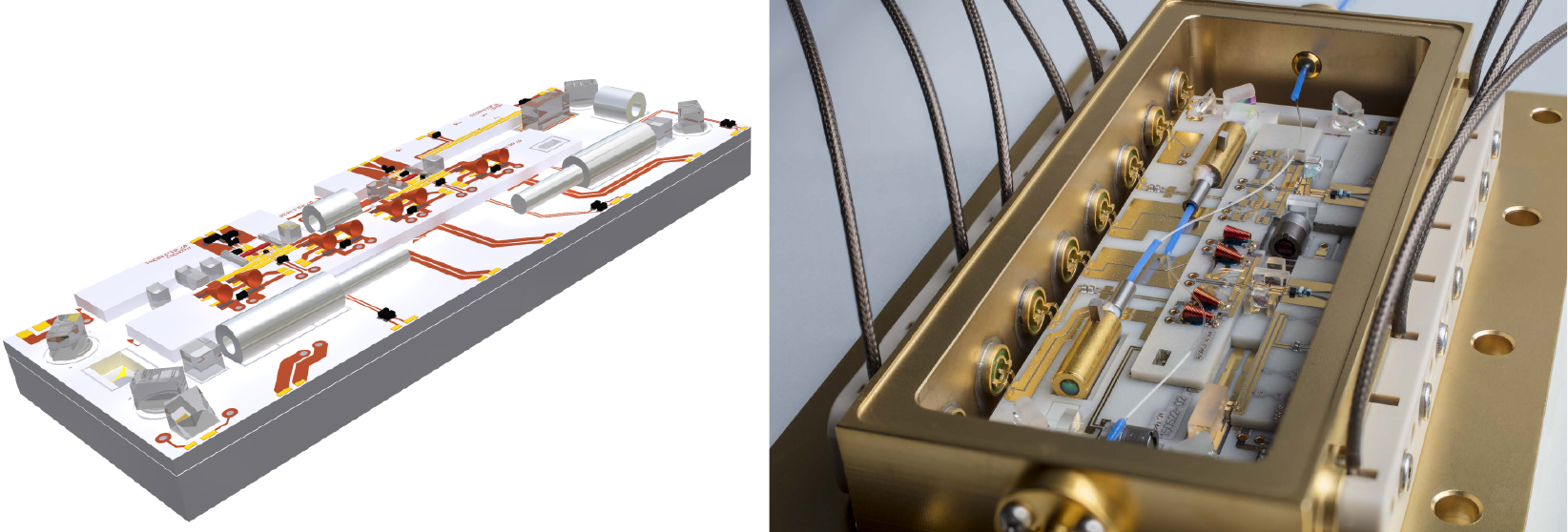}};
	
	\def\rend{-2.5}
	\def\lend{-4.5}
	\def\tend{1.8}
	\node[align=center] at (-6,\tend) {\textbf{a.}};
	\node[align=center, text=white] at (0.1,\tend) {\textbf{b.}};

	\draw[<-, line width=2pt] (-1.9,1.27)-- (-1.9,1.7) -- (\lend,1.7)  node[left]{VBHG};
	\draw[<-, line width=2pt] (-2.9,.75)-- (-2.9,1.34) -- (\lend,1.34)  node[left]{Chip 1};
	\draw[<-, line width=2pt] (-3.5,.5)-- (-3.5,.98) -- (\lend,.98)  node[left]{$\mu$-isolator};
    \draw[<-, line width=2pt] (-4.1,.18)-- (-4.1,.6) -- (\lend,.6)  node[left]{Chip 2};
    
    \draw[<-, line width=2pt] (-4.8,-1)-- (-4.8,-1.36) -- (\rend,-1.36)  node[right]{$\mu$-mirror};
    \draw[<-, line width=2pt] (-3.8,-.55)-- (-3.8,-1) -- (\rend,-1)  node[right]{Fiber Coupler};
	\draw[<-, line width=2pt] (-1.3,.75)-- (-1.3,-.8);

	\end{tikzpicture}
	
	\caption{\csentence{CAD and photo of ECDL-MOPA} \csentence{a)} CAD Model of an ECDL-MOPA Module with Dimensions 30 x 80 x 10 mm\textsuperscript{3}. VHBG: volume holographic Bragg grating; \textmu-isolator: micro-optical isolator; \textmu-mirror: micro-mirror. The optical fibers exiting the fiber coupler are not shown in the figure. \csentence{b)} ECDL-MOPA Laser Module for Operation at 1064 nm. The photograph shows the AlN ceramic body and the housing, as well as the optical and electrical feedthroughs.
	This figure is adapted from \cite{Wicht2017}.} \label{fig:LS_LS}
\end{figure}

As laser sources, semiconductor diode lasers are ideal candidates for BECCAL: they are compact, robust, available in a variety of wavelengths and expected to be resilient against the thermo-mechanical loads associated with launch and operation conditions in BECCAL~\cite{Wicht2017, Schiemangk2015,  Luvsandamdin2013, Luvsandamdin2014}. To this end, micro-integrated laser modules will be used within BECCAL, as in heritage missions~\cite{Schkolnik2017, Schkolnik2016, Pahl2019, Doringshoff2019, Dinkelaker2017}.
The laser modules are designed for versatility and multi-functionality through a design which incorporates two  semiconductor-based active or passive chips and two optical ports that can be used as input or output ports. 
For instance, a master-oscillator power-amplifier (MOPA) architecture can be implemented, with the master-oscillator consisting of either a monolithic diode laser chip, such as a distributed feedback laser, or as in the case of BECCAL, a discrete ECDL, thus creating an ECDL-MOPA. 
The laser chips are micro-integrated onto a micro-optical bench consisting of lithographically patterned aluminium nitride (AlN) substrates together with micro-optics and discrete (as opposed to integrated) electronics components. A CAD model of a MOPA with an ECDL as master-oscillator is shown in fig.~\ref{fig:LS_LS}a.

\begin{table}[t!]
	\caption{Main electro-optical expectations of the laser sources. PER: polarization Extinction Ratio; SMSR: Side Mode Suppression Ratio; FWHM: Full Width at Half Maximum.}
	\begin{tabular}{lcccc}
		\hline
		\\
		\textbf{Functionality}                    & \textbf{Rb}  & \textbf{K} & \textbf{Red-detuned}  & \textbf{Blue-detuned}  \\
		& & & \textbf{dipole trap} & \textbf{dipole trap} \\
		\\
		\hline
		\\
		Central wavelength {(}\si{\nano\meter}{)}      & \num{780.241}               & \num{766.701}               & \numrange{1054}{1074}            & $<$ \num{767}                  \\
		Output power {(}\si{\milli\watt}{)}            & \num{300}                   & \num{250}                    & \num{500}                     & \num{200}                      \\
		PER {(}dB{)}                     & -15                   & -15                    & -15                     & -15                      \\
		SMSR {(}dBc{)}                   & 30                    & 30                     & 30                      & 30                       \\
		linewidth (FWHM, \SI{1}{\milli\second}) {(}\si{\kilo\hertz}{)} & 100                   & 100                    & 100                     & 100       
		\\
		\\
		\hline
	\end{tabular} \label{tab:lsls}
\end{table}

The micro-optical bench is fully packaged into a housing made of Kovar, which is a commercial available nickel–cobalt ferrous alloy. 
The electrical signals are fed through via Mini-SMP coaxial connectors and the optical signals are fed through via single mode, polarization-maintaining optical fibers. 
In the ECDL-MOPA configuration, two optical output ports are available -- a main port that delivers the laser power to the physics package, and an auxiliary port, that can be used for the generation of beat notes or for monitoring purposes. 
The module housing has dimensions of \SI{125}{\milli\meter} x \SI{75}{\milli\meter} x \SI{22.5}{\milli\meter} and a mass of approximately \SI{760}{\gram}. 
Fig.~\ref{fig:LS_LS}b shows a picture of an ECDL-MOPA laser module. 
Such a laser module operating at a wavelength around \SI{1064}{\nano\meter} \cite{Kuerbis2019} has successfully been flown as part of an iodine-based frequency reference in the JOKARUS mission~\cite{Schkolnik2017, Doringshoff2019} on the TEXUS 54 sounding rocket in May 2018. 
It delivered more than \SI{500}{\milli\watt} output power ex-fiber within a technical (FWHM) linewidth of \SI{25}{\kilo\hertz} measured on a timescale of \SI{1}{\milli\second} (estimated from the frequency noise PSD according to the method proposed in~\cite{DiDomenico2010}) and a Lorentzian linewidth of smaller than \SI{1}{\kilo\hertz}. 
The ECDL-MOPA technology has already been transferred to the wavelength of \SI{780}{\nano\meter} for Rb-based experiments, such as MAIUS-B. 
At an emission wavelength of \SI{780.24}{\nano\meter} (corresponding to the Rb D${}_2$ line), an optical output power of the ECDL-MOPA of $\geq$\SI{400}{\milli\watt} has been demonstrated. 
Identical diode laser chips and micro-optics technologies can be used to produce the ECDL-MOPAs emitting at \SI{767}{\nano\meter} and for the blue-detuned optical trap. 
The expected performance of the lasers is shown in tab.~\ref{tab:lsls}.

\subsubsection{Free-space optics} \label{sec:ZD}

As outlined previously, the majority of light control is performed via free-space optics. However, the conditions during a rocket launch and aboard the ISS necessitate a light distribution system of very high thermal and mechanical stability as well as small size. In order to meet these requirements, we utilize a range of miniaturized optical benches based on a toolkit first introduced in~\cite{Duncker2014}.
This technology has successfully been used in previous sounding rocket missions~\cite{Schkolnik2016,  Dinkelaker2017, Lezius2016} and is going to be utilized within the MAIUS-2/3 missions~\cite{Mihm2019}.\\
The foundations of this toolkit are optical benches made from Zerodur, a glass ceramic produced and manufactured by Schott AG. This material has a near zero coefficient of thermal expansion and mechanical properties similar to those of aluminium. Onto these optical benches, free-space optical components such as mirrors, beam splitters, etc.\ are glued. By using the appropriate components, any functionality that can be achieved with a laboratory based optical setup can also be realized using these benches.

\begin{figure}

    \begin{tikzpicture}[>=stealth]
      \tikzset{every node}=[font=\footnotesize\sffamily]
      \node[inner sep=0pt] (Picture) at (0,0)
      {\includegraphics[width=0.95\textwidth]{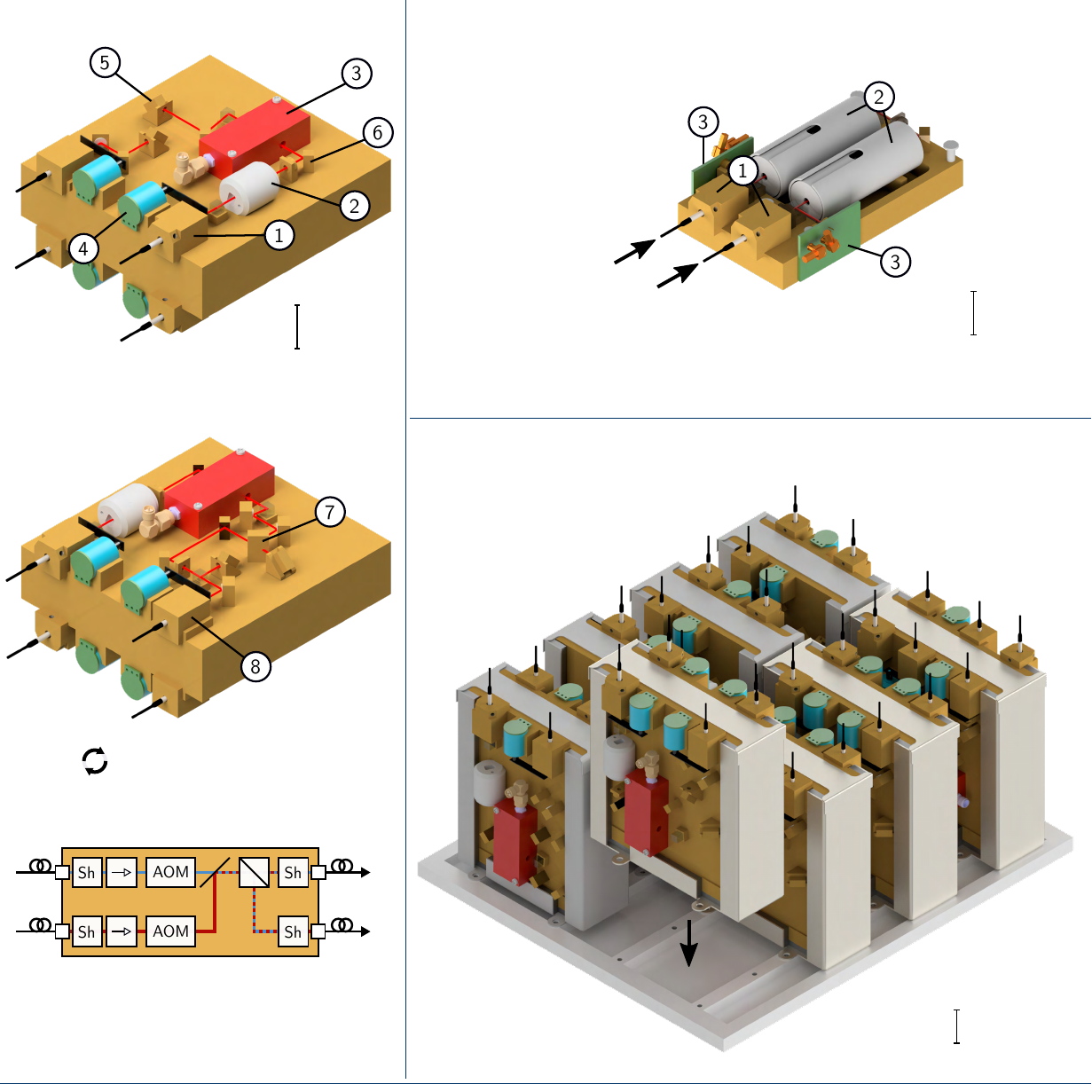}};
      \node[anchor=center] at (-4,-5.2) {Schematic drawing};
      \node[anchor=center] at (-4,-2.35) {Bottom side};
      \node[anchor=center] at (-4,1.9) {Front side};
      \node[anchor=north west] at (-1.4,1.2) {\textbf{c.} Optical bench rack and enclosure};
      \node[anchor=north west] at (-1.4,6.0) {\textbf{b.} Spectroscopy bench};
      \node[anchor=north west] at (-6.0,6.0) {\textbf{a.} Zerodur optical bench};
      \node[anchor=center] at (1.5,2.5) {Unmodulated beam};
      \node[anchor=center] at (0,3.7) {Modulated beam};
      \node[anchor=west] at (4.55,-5.32) {25 mm};
      \node[anchor=west] at (4.75,2.6) {25 mm};
      \node[anchor=west] at (-2.75,2.4) {25 mm};
    \end{tikzpicture}
	
	\caption{\csentence{a. top and center: Rendering of Zerodur optical bench} with: (1) fiber collimator, (2) optical isolator, (3) AOM, (4) shutter, (5) 45$^{\circ}$-mirror, (6) mirror, (7) dichroic mirror, (8) fiber coupler \csentence{a. bottom:} Schematic depiction of the same optical bench.
	\csentence{b. Rendering of optical bench for spectroscopy} with (1) fiber collimators, (2) spectroscopy cells with shielding, (3) photodiodes and PCBs for signal processing.
	\csentence{c. Rendering of optical bench mounting} The mounting structure for the optical bench safely locks the bench into place without applying any localized force that can lead to stress or deformation of the bench. The eight optical benches for light manipulation are grouped together in a standardized, modular system.
} \label{fig:LS_ZD}
\end{figure}
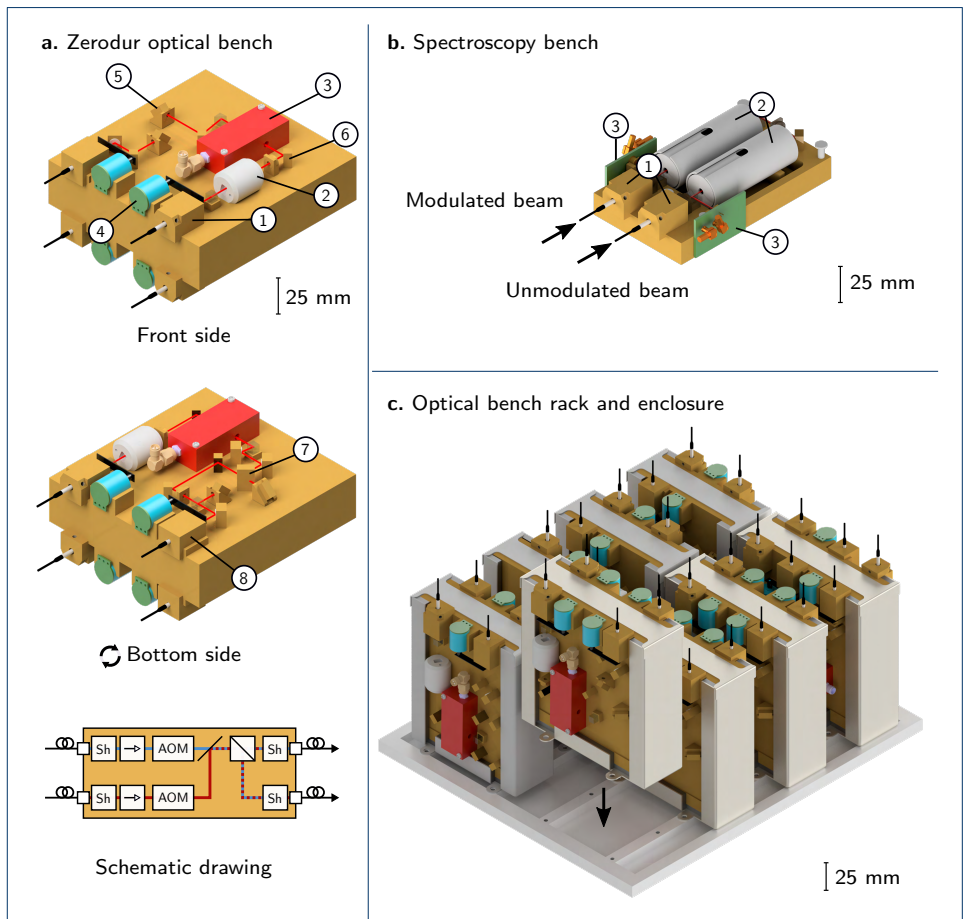

A total of ten optical benches will be used for BECCAL. Eight of those benches will be used for light distribution and preparation, and the remaining two as spectroscopy benches. 
A rendering of one of the optical benches used for light distribution can be seen in fig.~\ref{fig:LS_ZD}a: 
In a first step, the light from a fiber is collimated using a fiber collimator~(1). The light then passes an optical isolator~(2) to suppress back-reflections into the laser. Most experimental sequences require intensity control and fast switching. To achieve this, we use a conjunction of an AOM~(3) for fast suppression and a shutter~(4) for complete suppression. Mirrors~(6) are used to redirect the beam, and mirrors~(5) angled at 45$^{\circ}$ with respect to the bench guide light through a hole in the optical bench to the opposite side. Using these, we can use both sides of the optical bench, thereby effectively doubling the usable surface as compared to single-sided benches.
A dichroic mirror~(7) is used to overlap the beams at \SI{767}{\nano\meter} and \SI{780}{\nano\meter} in the same polarization state. We split a beam into two beams by using waveplates and polarizing beam splitters. The light is then coupled back into a fiber using a fiber-coupler~(8). We regularly achieve high fiber-coupling efficiencies of about \SI{81}{\percent}~\cite{Duncker2014}.

To match the stringent size and mass constraints aboard the ISS and to further improve stability, we have adapted and improved the above toolkit as follows~\cite{Marburger2019HighlyEnvironments}:
The benches are mounted in a clamp-like support structure that holds the bench from all sides (see fig.~\ref{fig:LS_ZD}c). Between the bench and the support, a rubber material is used as a cushion to dampen vibrations and mediate any direct force that could result in coupling-losses. The eight distribution benches are aggregated into a mounting structure as shown in fig.~\ref{fig:LS_ZD}c. This modular system uses a standardized optical bench form factor with \SI{30}{\milli\meter} thickness, \SI{125}{\milli\meter} length and \SI{100}{\milli\meter} or \SI{120}{\milli\meter} width. This enables easy exchange of benches on ground during implementation.

For frequency stabilization of the lasers, two spectroscopy modules are being used. A rendering can be seen in fig.~\ref{fig:LS_ZD}b: Two light beams, one modulated, one unmodulated, are coupled onto a bench using two fiber couplers~(1). They then counter-propagate through two spectroscopy cells~(2), filled with the relevant atomic gas. After passing the cells, the modulated and unmodulated beam are then each redirected onto a fast photodiode, the signal of which is further processed on a PCB~(3).

\subsubsection{Fiber-based optics} \label{sec:LS-fiber}

Before and after the free-space optical benches, light is controlled and manipulated in a fiber optic system. The fiber optic system comprises of fiber splitting/overlapping, switches, in-line photodiodes and optical modulators.

Several arrays of fiber splitters are used both to split single outputs of the free-space benches or lasers, and for combining multiple light paths. 
Fiber splitters are used in places where symmetric splitting is required.

Fiber switches will be used both in the detection and interferometry paths to provide the flexibility needed within the system.
Additionally to the switching capabilities provided by the AOMs and mechanical shutters included on the Zerodur free-space optical benches, the fiber switches also provide simultaneous extinction of multiple frequency components.

Radio-frequency-based fiber optics are used in the form of electro-optical modulators (EOMs) and one AOM.
Two EOMs add frequency side-bands needed for spectroscopy.
The AOM controls the intensity of the light for the first interferometry axis up to an extinction of \SI{-30}{\decibel}.

In-line photodiodes will be an essential part of the BECCAL monitoring concept. 
A large amount of house-keeping and control data will be taken to monitor and characterize the performance of the system during its lifetime. 
Optical taps (of typical ratio 99:1) provide an efficient way to monitor optical powers across the payload and provide both housekeeping and troubleshooting data if necessary.

\subsection{Control electronics}\label{sec:Electronics}
The control electronics system is the backbone of BECCAL, as it is in all quantum mechanics experiment. 
All active elements in BECCAL, which excludes metal parts and structural elements, are powered and controlled by the electronics presented in the following section.
The whole BECCAL apparatus will have approximately \SI{1.5}{\square\meter} of circuit board, the majority of which is custom designed.

\subsubsection{Power budget}
The power budget is governed by four main states: cold and hot standby, capacitor charging mode, and standard operational mode.
In cold standby, all non-vital components are turned off and only the vacuum pumps, the communication and the monitoring stay on. 
In this mode, the apparatus draws \SI{144}{\watt} of power. 
In hot standby, the system thermalizes prior to full operation mode. 
Here the laser system is switched on, with the rest of the apparatus in the same state as cold-standby.
This mode results in \SI{1066}{\watt} power consumption. 
In capacitor charging mode, the peak power increases to \SI{1467}{\watt}. 
The peak power takes into account the maximum magnetic field of \SI{175}{G} and all other systems in full operation.
During the actual operation of the experiment, the power consumption drops back to the \SI{1066}{\watt} of the hot standby mode because the chip and coil current drivers are supplied from capacitors and not from the \SI{28}{\volt} line.
In the standard operational mode, the apparatus alternates between charging the capacitors and running the experimental sequences. 
On average it will draw \SI{1267}{\watt} of power, which directly translates into heat that needs to be dissipated.
More detailed information on the distribution of the electrical power between the lockers can be found in tab.~\ref{tab:ELinputpower}.
These numbers represent a worst case with every device powered-up.
Depending on the actual experiment sequence, the real power will be lower, as unused parts can be powered down.

Due to the use of capacitors to store energy, the peak electrical power and the peak heat dissipation do not occur at the same time. 
Due to the cycle times of the experiments of typically a few seconds, the average power drives the thermal design (sec.~\ref{sec:SWaP}).
The electronic design, however, is driven by the peak electrical power.
Tab.~\ref{tab:ELheatdissipation} contains more detailed information on the heat dissipation per locker.

The electric power is drawn from four \SI{28}{\volt}-EXPRESS Rack power outlets as each one is individually limited to \SI{20}{\ampere} at nominal \SI{28}{\volt}.
This divides BECCAL logically in four groups: control electronics, chip\&coil current drivers and the rubidium and potassium laser electronics.
There is no power sharing between those groups.
The power supply of the titanium sublimation pump is part of the chip\&coil current driver group.
In order not to trip the circuit breaker of the EXPRESS rack, there are additional electronic circuit breakers in BECCAL.

\subsubsection{Current drivers}
In order to ensure the lowest possible noise for currents through the coils and the atom chip, the corresponding current drivers are powered by supercapacitors with output voltages between \SI{5}{\volt} and \SI{48}{\volt}.
The current drivers themselves use a linear design to avoid switching noise.
There will be separate electronic circuit breakers to protect the atom chip against faults in the current drivers.

The laser current drivers also use a linear design, with additional filtering in order to suppress power supply noise.

\subsubsection{Experimental control, timing and data storage}
All frequency and timing related signals are phase locked to a central oven-controlled cristal oscillator (OCXO). 
This ensures that the timing of all events are synchronized and all frequencies are as precise as the central reference. 
There will be a slow tracking of the OCXO frequency against GPS via the network time service of the ISS for post-corrections.
All RF signal generators are direct digital synthesizers (DDS) based and generate signals in the range of \SIrange{0.3}{150}{\mega\hertz} with a frequency resolution better than \SI{0.1}{\hertz} and will be capable of ramps and jumps in frequency, phase, and amplitude simultaneously.
The microwave generators operate between \SI{250}{\mega\hertz} and \SI{6.8}{\giga\hertz} and are phase locked to DDS signal generators.

The main computer that controls the experiment will have a dual core Intel I7 processor with \SI{8}{\giga\byte} RAM and \SI{500}{\giga\byte} of non-volatile storage on two internal SSDs and another \SI{500}{\giga\byte} on two external SSDs for experimental data.

The computer communicates to the ISS-Network and to the internal network via Ethernet.
The internal network is based around a custom protocol on plastic optical fibers (POF), combining data, frequency, and trigger signals in a single wire pair, saving a lot of weight and space while at the same time avoiding ground loops.

\begin{table}[t!]
	\caption{Electrical input power per wall plug: control electronics (CE), chip\&coil current and titanium sublimation pump (CCC/TSP), rubidium laser system (LS-Rb), and potassium laser system (LS-K). Due to the usage of capacitors, the electrical peak power consumption does not occur in the same experimental phase as the peak heat dissipation listed in tab.~\ref{tab:ELheatdissipation}.}
	\begin{tabular}{lSSSSS}
		\hline
		\\
		\textbf{State} & \textbf{CE} & \textbf{CCC/TSP} & \textbf{LS-Rb} & \textbf{LS-K} & \textbf{Total power} \\
		 & \textbf{(W)} & \textbf{(W)} & \textbf{(W)} & \textbf{(W)} & \textbf{(W)} \\
		\\
		\hline
		\\
		Cold standby               & 144 &   0 &   0 &   0 &  144 \\
		Hot standby                & 324 &  84 & 318 & 340 & 1066 \\
		Capacitor charging         & 324 & 485 & 318 & 340 & 1467 \\
		Experiment operation       & 324 &  84 & 318 & 340 & 1066 \\
		Avg. charging + operation  & 324 & 285 & 318 & 340 & 1267 \\
		Ti-sublimation pump        & 144 & 372 &   0 &   0 &  516 \\
		\\
		\hline
	\end{tabular}\label{tab:ELinputpower}
\end{table} 

\begin{table}[t!]
	\caption{Heat dissipation per locker: control electronics (CE), physics package (PP), and laser system (LS). Due to the usage of capacitors, the electrical peak power consumption, shown in tab.~\ref{tab:ELinputpower}, does not occur in the same experimental phase as the peak heat dissipation.}
	\begin{tabular}{lSSSS}
		\hline
		\\
		\textbf{State} & \textbf{CE (W)} & \textbf{PP (W)} & \textbf{LS (W)} & \textbf{Total heat (W)} \\
		\\
		\hline
		\\
		Cold standby               & 123 &  21 &   0 &  144 \\
		Hot standby                & 331 &  77 & 658 & 1066 \\
		Capacitor charging         & 367 &  77 & 658 & 1102 \\
		Experiment operation       & 586 & 187 & 658 & 1431 \\
		Avg. charging + operation  & 477 & 132 & 658 & 1267 \\
		Ti-sublimation pump        & 123 & 393 &   0 &  516 \\
		\\
		\hline
	\end{tabular}\label{tab:ELheatdissipation}
\end{table}

\subsection{Software operation}\label{sec:SW}

BECCAL’s operation will be conducted in collaboration with scientists from many different institutions. 
The main goal is to enable all participants to have their experiments executed on the BECCAL instrument.
This requires a robust structure and an easy-to-use software interface for the scientists.

For successful operation, four main software tools have been identified:
the Experiment Control Software (ECS), the Experiment Design Tools (EDTs), the Lab Operation Software (LOS), and the Ground Control Software (GCS).
All of the mentioned software tools use model-driven software technologies to either support users with the creation of valid experiment definitions, or to generate source code from engineering data of the experiment hardware.

\subsubsection{Experiment control software}

The ECS is the software running on the on-board computer of the experiment. 
It communicates directly with the experiment electronics and is in charge of executing the experiment step by step.
It is controlled through the LOS or the GCS depending if it runs on one of the testbeds or the BECCAL instrument on-board the ISS.
The drivers for the control electronics described in sec.~\ref{sec:Electronics} are largely generated from a model description.
The capabilities of every electronic card and the stack it resides in are captured using a domain specific language~\cite{weps2018}. 
Those descriptions serve as an unambiguous single point of truth and provide an interface between the electronic and software development domain. 
Changes in the firmware of an electronics card can then be reflected in the ECS with very little manual effort which saves resources during the development of an experimental setup such as BECCAL.

The core elements for experiment execution are sequences and subsequences. 
Each sequence captures the next actions for the electronics in a precisely timed schedule in order to manipulate the state of the experiment.
Frequently occurring actions can be placed in a subsequence, sharing common actions among sequences and avoiding code duplication.

A set of sequences is then assembled into the experiment-execution graph (EEG), which, besides serial execution, also provides mechanisms to direct the experiment flow~\cite{weps2018}.
In this way the experiment can, for example, run through certain sequences multiple times with different input parameters and tune the experiment before entering into the next phase without the manual interaction of an operator.

This general separation between sequences and experiment-execution graph has already been used for the MAIUS-1 mission~\cite{Becker2018}.
There, from the EEG, sequences, and subsequences, the corresponding C++ code was generated and compiled as part of the ECS. 
For BECCAL this approach is not feasible as a wider range of experiments needs to be supported in a transparent and efficient way.
Therefore, the graphs are stored as separate files.
The ECS for BECCAL includes a newly developed graph-interpreter engine which interprets and executes the graphs directly from those files after successful sanity checks have been carried out.
To ensure safety of the experiment at all times only EEGs which have been previously qualified on a ground test bench will be uploaded to the ISS through the ground-control station.
After experiment execution the full set of captured scientific data is stored on the on-board computer’s hard drive and has to be downloaded before it can be distributed to the author of the experiment.

\subsubsection{Experiment design tools}

The EDTs are the software components for the participating scientists in order to create a fully functional experiment.
They support the scientists with the task of creating a formally correct experiment which can then be tested on one of the ground-based testbeds and later be qualified for the execution on board the ISS. 
For this purpose, the sequences and subsequences, as well as the EEG, are also created using a model representation in the background.
To this end a domain specific language (DSL) was developed which describes sequences and subsequences in a human- and machine-readable textual representation, allowing for easy integration into version control systems.
Graphical user interfaces (GUIs) allow experimenters to create and edit these sequences, as well as the EEGs, in a convenient way.
Both interfaces facilitate complex validation procedures using the model description of the hardware to indicate errors and potential problems to the user via visual markers.
This model-driven approach also restricts the allowed features for the scientist to a set which is qualifiable for the execution on board the ISS.

\subsubsection{Ground control software}

\begin{figure}
	\includegraphics[width=0.9\textwidth]{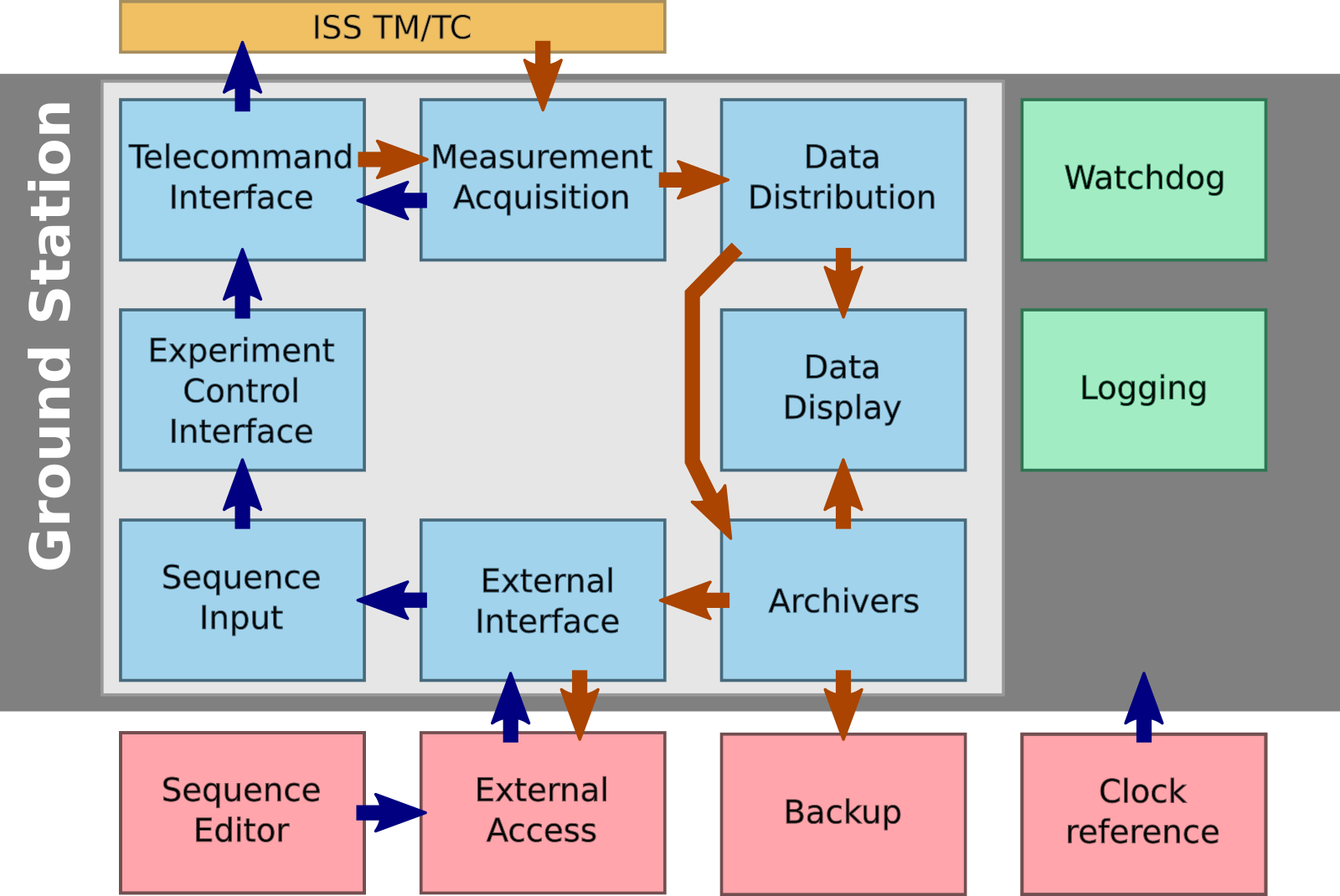}
	\caption{\csentence{BECCAL Ground Control Software overview}
	 Several software modules enable telecommanding (TC) and telemetry (TM) for controlling and monitoring the instrument.
	 Inside the ground station (gray box) data are received from and sent to the ISS.
	 Display modules enable the ground control personell to supervise the experiment status and scientific data.
	 All data are archived and stored in off-site backups.
	 An external interface is provided to allow upload of new sequences and download of data for scientists.
	 Logging and watchdog facilities ascertain that all essential software modules are always running.}
	\label{fig:SW-gsw}
 
\end{figure}

\noindent
The GCS is the single point of communication to the BECCAL instrument on board of the ISS during normal operation.
It comprises several software modules, which are depicted in fig.~\ref{fig:SW-gsw}.
A telecommand interface provides means to control the experiments, while a telemetry interface receives live housekeeping data at all times and scientific data during operation.
Custom protocols for telecommands and telemetry will be used alongside standard tools such as \textit{ssh} and \textit{rsync} for maintenance and file transfers.
Data is distributed by the telemetry module via network to display modules for operators to supervise operation and to a central archive, which is copied to one or more off-site backups.
In the archive and the backups, redundant file systems with checksumming-capabilities will ensure data integrity.
From there, a web interface allows scientists to access data of their experiments.
The same interface will allow scientists to upload new experimental sequences and EEGs, which are then validated and uploaded to the ECS.
A logging service keeps track of all events and inputs for debugging purposes, and a watchdog services ensures that all critical software modules, are running at all times.

The most important objective is the safe operation of the instrument at all times.
Therefore, only qualified sequences and EEGs are allowed to be uploaded to and executed on the instrument.
The GCS is responsible for ensuring data integrity of the uploaded data before execution. 
It also collects and monitors BECCAL’s telemetry data and executes standard maintenance tasks.
If non-nominal telemetry is received, the GCS is also capable of aborting ongoing experiments if necessary, and bringing the instrument back in a nominal state.
After an experiment execution is finished, the GCS initializes the download of all generated scientific data from the experiment which can subsequently be distributed to the responsible scientists.

\subsubsection{Lab operation software}

The LOS is responsible for communication with the ECS in the laboratory environment of the testbed.
It commands the ECS to execute EEGs or sequences and presents the received live telemetry data.
Compared to the GCS it facilitates a more direct interaction with the ECS, allowing, for example, the execution of not yet qualified sequences or graphs. This enables the operators to debug or optimize parts of an experiment.
The LOS operates on the same model representation as generated by the EDTs, anomalies which occur during execution can therefore directly be discussed with the author of the experiment with the same unambiguous input data.

\section{Operations}
\label{sec:OPs}

BECCAL will be installed in the US Destiny module on board the ISS. 
Operations include the delivery to the launch site, launch, installation in orbit, check-out procedures, operations, data distribution, proprietary rights, removal from orbit, and post-flight operations. 

\subsection{Launch}

After integration in Germany, the payload is delivered to the designated launch site, where it is integrated into the transport capsule. 
At the time of writing, the exact launcher has not been confirmed, and as such, details of related operations and requirements will be decided upon once the launch vehicle is known. 
The current design allows for transport with any of the currently available launchers.

Independent of the exact launcher, the experiment is expected to be exposed to vibrations of \SI{8.8}{g_{rms}} and shocks of up to \SI{12}{g} during launch.
Therefore the hardware must be tested prior to flight to ensure it is able to withstand the expected launch. 

\subsection{Installation and check-out}

On arrival in orbit, the payload will be installed by the crew, which includes establishing interfaces to the ISS and between lockers. 
Installation is completed with the initial power-up of the payload which is done by both the crew in orbit and the crew on ground. 

After installation, BECCAL will undergo a series of test procedures to assure operation readiness and to verify remaining requirements (such as those only achievable in microgravity). 
This process includes the check of the health and status data as well as the execution of the first experimental sequences. 

\subsection{On-orbit operation}

Following successful payload check-out, BECCAL will go into nominal operation. 
In nominal operation, BECCAL will execute experimental sequences, which are prepared on ground and tested using a ground testbed. 

During operation, the experiment will be controlled and monitored via a datalink from ground (sec.~\ref{sec:SW}).
The quiet times onboard the ISS are preferred for operation to avoid disturbances by excess vibrations due to astronaut activities.

The operation of BECCAL is divided into three parts: cold standby, hot standby, and experimental operation.
The later one can be further divided into charging and experimental phases.
Technical details are discussed in sec.~\ref{sec:Electronics}

\paragraph*{Cold standby}
During cold standby, all non-vital components are off. Such that the ion getter pump is operational, and the payload computer records health and status data.

\paragraph*{Hot standby}
Approximately thirty to sixty minutes prior to experimental operation, the payload is set into hot standby. 
During hot standby the subsystems are prepared for experimental operation. 
In particular, this includes the thermalization of the laser system. 
After experimental operation, the system is kept in hot standby for performance checks before it is send to cold standby. 

\paragraph*{Experimental operation}
Experimental operation is divided into charging and experimental phases. 
During the experimental phase, the pre-programmed and tested sequences are executed. 
The two phases are of roughly equal duration. 

The duration of experimental operation is based on the rhythm of the crew on board and the available resources both on ground and in orbit. 

\subsection{Performing an experiment on BECCAL}
\begin{figure}
	\includegraphics[width=0.9\textwidth]{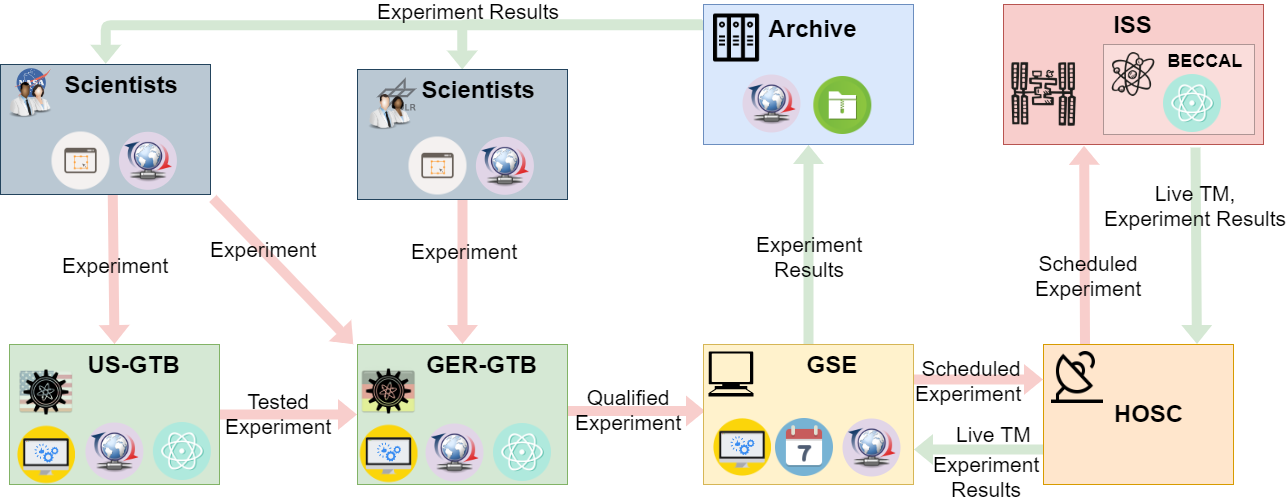}
	\caption{\csentence{BECCAL operation and relevant information flow} Basic flow of experiment data for the operation of BECCAL. Prepared experiments are passed by the scientists to a GTB for qualification before being uploaded and executed on board the ISS via the GSE (red arrows). The scientific results are downloaded from the ISS and distributed to the respective authors (green arrows). The data is routed through the Huntsville Operations Support Center (HOSC).}

	\label{fig:SW-data-flow}
\end{figure}

The process of performing an experiment on BECCAL has several stages: 
First, participating scientists, who may not have direct access to one of the ground-based testbeds, need  to design formally correct experiments for BECCAL which they then distribute to one of the testbeds for execution.
Testbed operators take those experiment designs, check them for correctness, execute them on the testbed, and return the experiment results to the respective authors.
If the experiment is deemed ready for execution on the BECCAL instrument, it is sent to the German ground-based test bench for qualification.
Once an experiment is successfully qualified no further changes are possible without re-qualification.
The qualified experiment is finally forwarded to the BECCAL ground-control station and scheduled for on-board operation. 
Upon download of the experiment's scientific data set the results are distributed to the authors.
Fig.~\ref{fig:SW-data-flow} presents a high-level overview of the different stages and the information flow in between.

\subsection{Maintenance}

BECCAL will make use of orbital replaceable units in order to provide some degree of redundancy in case of an error that cannot be resolved from ground.
In such cases, each enclosed locker structure (laser system, physics package, control electronics) can be exchanged with a replica system already assembled and tested on ground. Additionally, the lasers modules are themselves housed in smaller ORUs such that they can replaced without replacing the entire laser locker as discussed in sec.~\ref{sec:LS}.

\subsection{Lifetime and de-commissioning}
BECCAL shall be operational for a minimum of twelve months and is intended to be installed in orbit for three years. 

After the in-orbit period of three years, the condition of the payload and the available resources will be assessed before BECCAL is de-integrated. It is not planned for the hardware to returned to ground. 

\section{Summary of experimental capabilities}\label{sec:Sum}
\begin{table}[ht!]
\caption{Overview of the experimental capabilities of BECCAL}
\begin{tabular}{p{0.5\textwidth}p{0.4\textwidth}}

  \hline	
   & \\
  \textbf{Atom numbers} & \\
  Single $^{87}$Rb BEC & $\geq$ \num{1e6} \\
  Single $^{41}$K BEC & $\geq$ \num{1e5}  \\
  Dual-species BEC of $^{87}$Rb and $^{41}$K (respectively)  & $\geq$ \num{1e5}, $\geq$ \num{1e4}   \\
  Trapped $^{85}$Rb atoms &  $\geq$ \num{1e6} \\
  Trapped $^{39}$K atoms &  $\geq$ \num{1e5} \\
  Trapped $^{40}$K atoms &  $\geq$ \num{1e4} \\
  & \\
  \textbf{Atom trapping and manipulation} & \\
  Magnetic traps: available structures & Atom chip and four pairs of coils \\
  Crossed optical dipole trap: wavelength, beam waist, total power & \SI{1064(10)}{\nano \meter}, \SI{100(20)}{\micro \meter}, \SI{300}{\milli\watt} \\
  Optical lattice & Switchable retro-reflection of the beams of the crossed optical dipole trap \\
  Painted potentials: barrier height, sizes of rings, total power & \SI{5}{\micro \kelvin} with contour length $\leq$\SI{100}{\micro \meter} and refresh rate $\geq$ \SI{100}{\hertz}, inner diameter \SIrange{20}{1500}{\micro \meter} and outer diameter \SIrange{40}{2000}{\micro \meter},  \SI{50}{\milli\watt} optical power at \SI{764}{\nano\meter}\\
  Life times of single species clouds: non-trapped, magnetically, and optically trapped (respectively)  & $\geq$ \SI{10}{\second}, $\geq$ \SI{3}{\second}, $\geq$ \SI{10}{\second} \\  
  Feshbach-field & $\geq$ \SI{175}{G} with tunable modulation of up to \SI{1}{G} at $\geq$ \SI{0.1}{\kilo Hz} \\
  Transport of atoms away from the chip & \SIrange{0.1}{2}{\milli \meter} \\
  Minimum expansion velocity for $^{87}$Rb,  $^{41}$K, and $^{87}$Rb-$^{41}$K mixtures after atomic lensing & $\leq$ \SI{100}{\micro \meter \per\second} \\
  Two independent radio-frequency sources & \SIrange{0.1}{25}{\mega \hertz} with Rabi frequency $\geq$ \SI{10}{\kilo \hertz} and ramp rates $\geq$ \SI{100}{\mega\hertz\per\second}\\ 
  Microwave generator & \SI{6.835}{\giga\hertz}, \SI{3.036}{\giga\hertz}, \SI{1285}{\mega\hertz}, \SI{462}{\mega\hertz} and \SI{254}{\mega\hertz}  adjustable with \SI{30}{\mega \hertz} around the central frequency, Rabi frequency $\geq$ \SI{10}{\kilo \hertz} and ramp rates $\geq$ \SI{100}{\mega\hertz\per\second} \\
   & \\
  \textbf{Atom interferometry, two independent axes} & \\
  \textbf{Primary axis} &  parallel to Earth’s acceleration (nadir) and parallel to the chip surface \\
  Total free evolution time & $\geq 2T = \SI{2.6}{\second}$\\
  Raman diffraction beams & splitting efficiency $\geq$ \SI{90}{\percent}, $1/e^2$ beam diameter $\geq$ \SI{6}{\milli\meter}, Rabi frequency \SIrange{1}{50}{\kilo \hertz}, detuning adjustable \SIrange{1}{5}{\giga \hertz}, homogeneous magnetic field of \SIrange{0.05}{1}{G}, power of \SI{15}{\milli\watt} (\SI{70}{\milli\watt}) per frequency component in primary axis (secondary axis)\\
  Rotation compensation & retro-reflection mirror on tip-tilt stage for rotation compensation of ISS for times $\geq 2T = \SI{2.6}{\second}$ ($\geq$ \SI{3}{\milli \radian}) \\
  Coupling with external sensors & rotation sensor with noise floor \mbox{$\leq$ \SI{50}{\micro\radian\per\second\per$\sqrt{\mathrm{\hertz}}$}}\\
  \textbf{Secondary axis} & perpendicular to the orbital plane and perpendicular to the chip surface \\
  & \\
  \multicolumn{2}{l}{\textbf{Two orthogonal, species selective, spatially resolved absorption detection systems}} \\
  Spatial resolution & $\leq$\SI{10}{\micro \meter} \\ 
  Field of view & $\geq$\SI{10}{\milli\meter\squared} \\
  Scanning range of focal plane &  $\geq$\SI{10}{\milli\meter},\\
  Adjustable depth of field & \SIrange{100}{2000}{\micro\meter} \\ 
  & \\
  \textbf{Fluorescence detection} & single axis, species selective \\
  \\
  \textbf{Miscellaneous} & \\
  Multiple surfaces with different electrical properties on the atom chip & size $\geq$\SI{1}{\milli\meter\squared}\\
   & \\
  \hline
\end{tabular}
  \label{tab:OverviewTabel}
\end{table}

BECCAL is expected to offer a high flux of ultracold rubidium and potassium gases.
These gases can be brought to quantum degeneracy in less than \SI{2}{\second}, supporting experiments that require extensive statistics. 
The magnetic fields are generated by a multi-layer atom chip, with a design similar to the ones used in QUANTUS~\cite{Rudolph2015} and MAIUS~\cite{Becker2018}, in combination with four pairs of coils, three of which are oriented perpendicular to each other.
The fourth pair of coils produces a large magnetic field to tune the scattering length of the atoms. 
Additionally, the atoms can be manipulated with radio frequency and microwave fields.
This setup allows the creation of adiabatic-dressed potentials for hollow BECs or the formation of an atom laser. 

Furthermore, the atoms can be exposed to blue-detuned laser light from two perpendicular directions.
These beams can be spatially controlled to create traps of arbitrary shapes in three dimensions. 
Red-detuned light can also be used for crossed dipole trapping or the creation of optical lattices.

The atom chip will have different areas with metallic coating for investigations of atom-surface effects. 
The detection is carried out by either capturing the fluorescence signal of the atoms or by taking spatially resolved absorption images.
The absorption imaging systems will allow for changing the focal plane and the field of view. 

Moreover, BECCAL will have two perpendicular atom interferometry axes. 
One axis is nadir pointing and so is in the direction of the gravitational acceleration. 
Since this configuration is prone to contrast losses due to the rotation of the ISS, the retro-reflection mirror is mounted on a piezo-tip-tilt stage inside the vacuum chamber.
An overview of the core parameters is given in tab.~\ref{tab:OverviewTabel}.

\section{Conclusion}\label{sec:Conclusion}
We have presented the current design of an apparatus to perform experiments with ultracold and condensed atoms on board the International Space Station.
Despite the constraints on size, mass and power, the stringent safety requirements, and the requirements on the robustness, BECCAL is designed to support a broad range of experiments from fundamental physics to the development of quantum sensors. 

Driven by anticipated scientific investigations, and in line with the current design, we have reported the core features of the device, including the ability to generate ultracold or condensed ensembles of rubidium, potassium and mixtures of both by utilizing a variety of magnetic or optical traps as well as tools their coherent manipulation.
We expect BECCAL to advance the current understanding in the fields of quantum optics, atom optics, and atom interferometry in the unique conditions of microgravity linked to and beyond currently discussed ideas in the field.


\begin{backmatter}

\section*{Competing interests}
  The authors declare that they have no competing interests.

\section*{Author's contributions}
All authors contributed to the writing of the manuscript.

The systems engineering, thermal design and simulation, mechanical and interface design as well as the design of the vacuum / pump system of BECCAL has been surveyed by a group led by JG and CB. The work of the group has been distributed among JP, MW, and JG.

The laser subsystem has been led by VAH, with design work distributed between VAH (overall), AB (lasers), JPM, and AW (zerodur benches).  

The electronics are designed by TW and his group.

KF, SvAb, DB, CS, and WH designed the physics package supported by PF for coils and the magnetic shield.

MAE, MaMe, AR, and WPS provided theory support with MaMe coordinating the effort. 

NY, JK, SWC, and UI worked on the development of the realization of the blue box potential. 

The science definition team (SDT) consists of DMSK, BKS, NL, HM (US-SDT) and WPS, EMR, DB, MK (GER-SDT). 

\section*{Acknowledgements}
The herein described project is a bilateral collaboration between NASA and DLR, both contributing to the scientific and operational organization. 

This work is supported by the German Space Agency (DLR) with funds provided by the Federal Ministry for Economic Affairs and Energy (BMWi) due to an enactment of the German Bundestag under Grant Nos. DLR50WP1431-1435, 50WM1131-1137, 50MW0940, 50WM1240, 50WM1556, 50WP1700-1706, 50WP1806, 50WM1956, 50RK1957, by ``Nieders\"achsisches Vorab" through the ``Quantum- and Nano-Metrology (QUANOMET)" initiative within the project QT3, through the Deutsche Forschungsgemeinschaft (DFG, German Research Foundation) under Germany's Excellence Strategy – EXC 2123 QuantumFrontiers, Project-ID 390837967, and through ”Förderung von Wissenschaft und Technik in Forschung und Lehre” for the initial funding of research in the new DLR-SI Institut. MAE thanks the Center for Integrated Quantum Science and Technology (IQ$^{\text{ST}}$) for financial support.
The project was carried out in part at the Jet Propulsion Laboratory, California Institute of Technology, under a contract with the National Aeronautics and Space Administration (80NM0018D0004). 

The team acknowledges the contributions from NASA and their aid in adapting the payload to the needs of the International Space Station. We especially acknowledge the contributions from NASA Headquarters, Glenn Research Center, and Johnson Space Center. 
  

\bibliographystyle{bmc-mathphys} 
\bibliography{main}      





\end{backmatter}
\end{document}